\documentclass[a4paper]{article}

\usepackage{amssymb}
\usepackage{amsmath,amsthm}      
\usepackage{graphicx}            
\usepackage{microtype}           
\usepackage{booktabs}            
\usepackage{multicol}            
\usepackage[hidelinks]{hyperref} 
\usepackage{physics}
\usepackage{mathtools}
\usepackage{relsize}
\usepackage{setspace}
\usepackage{gensymb}

\usepackage{perpage} 
\usepackage{chngcntr}
\theoremstyle{definition}
\usepackage[titletoc,title]{appendix}
\usepackage{tocloft}
\usepackage[all]{xy}
\usepackage{sectsty}
\usepackage[top=25mm, bottom=25mm, left=15mm, right=15mm]{geometry}
\usepackage{authblk}
\usepackage[titletoc,title]{appendix}

\theoremstyle{definition}

\newtheorem{conjecture}{Conjecture}

\newcommand{\prt}[1]{\partial_{#1}}
\newcommand{\inv}[1]{\frac{1}{#1}}

\newcommand{\eps}{\epsilon}

\newcommand{\vx}{\vb{x}}

\newcommand{\mathsym}[1]{{}}
\newcommand{\unicode}[1]{{}}
\newcommand{\ba}{\begin{align}}
\newcommand{\ea}{\end{align}}
\DeclareMathOperator{\sinc}{sinc}
\DeclareMathOperator{\comb}{comb}
\DeclareMathOperator{\rect}{rect}
\newtheorem{theorem}{Theorem}
\newtheorem{proposition}{Proposition}
\newtheorem{lemma}{Lemma}

\usepackage{natbib}
\setcitestyle{super,open={},close={},comma}

\title{The continuum limit of k-space cavity angular momentum is controlled by an infinite range difference operator.}
\author[1]{Per Kristen Jakobsen}
\author[2]{Masud Mansuripur}
\affil[1]{Department of Mathematics and Statistics, the Arctic University of Norway,\linebreak Troms\o, Norway.}
\affil[2]{Wyant College of Optical Sciences, The University of Arizona, Tucson, USA. }

\begin{document}

\maketitle

\begin{abstract}
A wavepacket (electromagnetic or otherwise) within an isotropic and homogeneous space can be quantized on a regular lattice of discrete $\mathbf{k}$-vectors. Each $\mathbf{k}$-vector is associated with a temporal frequency $\omega$; together, $\mathbf{k}$ and $\omega$ represent a propagating plane-wave. While the total energy and total linear momentum of the packet can be readily apportioned among its individual plane-wave constituents, the same cannot be said about the packet’s total angular momentum. One can show, in the case of a reasonably smooth (i.e., continuous and differentiable) wave packet, that the overall angular momentum is expressible as an integral over the k-space continuum involving only the Fourier transform of the field and its $\mathbf{k}$-space gradients. In this sense, the angular momentum is a property not of individual plane-waves, but of plane-wave pairs that are adjacent neighbors in the space inhabited by the $\mathbf{k}$-vectors, and can be said to be localized in the $\mathbf{k}$-space. Strange as it might seem, this hallmark property of angular momentum does not automatically emerge from an analysis of a discretized $\mathbf{k}$-space. In fact, the discrete analysis shows the angular momentum to be distributed among $\mathbf{k}$-vectors that pair not only with nearby $\mathbf{k}$-vectors but also with those that are far away. The goal of the present paper is to resolve the discrepancy between the discrete calculations and those performed on the continuum, by establishing the conditions under which the highly non-local sum over plane-wave pairs in the discrete $\mathbf{k}$-space would approach the localized distribution of the angular momentum across the continuum of the $\mathbf{k}$-space.
\end{abstract}

\tableofcontents

\section{Introduction}
According to the classical theory of electrodynamics, an electromagnetic (EM) wavepacket propagating in free space carries energy, linear momentum, and angular momentum \cite{Cohen,Haus,Grynberg,Barnett2,Masud1,Marrucci,Masud2}. The knowledge of the wavepacket’s electric-field profile in spacetime, $\mathbf{E}(\mathbf{r},t)$, is generally all that is needed to compute its remaining physical properties\cite{Jackson}. The magnetic field distribution $\mathbf{H}(\mathbf{r},t)$ is readily computable from the (four-dimensional) Fourier transform of $\mathbf{E}(\mathbf{r},t)$, whence the Poynting vector $\mathbf{S}(\mathbf{r},t)=\mathbf{E}(\mathbf{r},t)\cross \mathbf{H}(\mathbf{r},t)$ yields the rate of flow of EM energy per unit area per unit time\cite{Jackson}. The linear momentum density is subsequently determined as $\mathbf{P}(\mathbf{r},t)=\mathbf{S}(\mathbf{r},t)/c^2$, and the angular momentum density as  $\mathbf{J}(\mathbf{r},t)=(\mathbf{r}\cross \mathbf{S}(\mathbf{r},t))⁄c^2$ , the latter encompassing both spin and orbital angular momenta of the wavepacket \cite{Cohen,Grynberg,Masud1,Masud2}.
Quantization of the EM field in free space entails the confinement of the field to a large, fictitious, $L\cross L\cross L$ cube within the Cartesian $xyz$ coordinate system, followed by imposition of periodic boundary conditions at the cube’s facets and evaluation of the field’s energy and linear and angular momenta contained within the cube in the limit when $L\rightarrow\infty$. In this scheme, the propagating plane-wave modes of the EM field are found to have $\mathbf{E}$-field amplitude $\mathbf{E}_0^{(n)}e^{i[k^{(n)}\cdot r - \omega^{(n)} t]}$  and H-field amplitude $\mathbf{H}_0^{(n)}e^{i[k^{(n)}\cdot r - \omega^{(n)} t]}$, where

\begin{align*}
\mathbf{k}^{(n)}=\frac{2\pi}{L}(n_x \hat{\mathbf{x}}+n_y \hat{\mathbf{y}}+n_z \hat{\mathbf{z}}),\;\;\ \omega^{(n)}=c\abs{\mathbf{k}^
{(n)}}, \;\; \mathbf{H}_0^{(n)}=\frac{\mathbf{k}^{(n)}\cross \mathbf{E}_0^{(n)}}{\mu_0 \omega^{(n)}}.
\end{align*}

Here, $n_x,n_y,n_z$ are positive, zero, or negative integers, $\omega^{(n)}\ge 0$ is the mode’s temporal frequency, $c=(\mu_0\varepsilon_0 )^{-\frac{1}{2}}$ is the speed of light in vacuum, and $\mu_0$ and $\varepsilon_0$ are the permeability and permittivity of free space, respectively. Each mode of the EM field can have a polarization state—i.e., linear, circular, or elliptical—specified by one or the other member of a pair of generally complex-valued, mutually orthogonal, unit-vectors $(\hat{e}_1,\hat{e}_2 )$, where
\begin{align*}
    \hat{e}_1\cdot\hat{e}_1^*=\hat{e}_2\cdot\hat{e}_2^*=1, \;\;\hat{e}_1\cdot\hat{e}_2^*=0,\;\; \hat{e}_1\cdot\mathbf{k}^{(n)}=\;\;\hat{e}_2\cdot\mathbf{k}^{(n)}=0.
\end{align*}

The total energy content of a wavepacket consisting of a superposition of discrete plane-waves can be shown to be the sum of the energies of the individual modes. Similarly, the linear momentum of the packet can be expressed as the sum of the linear momenta of its individual plane-wave constituents. However, these individual (discrete) plane-waves do not carry angular momentum (neither spin nor orbital); the overall angular momentum of the packet arises from interference between distinct pairs of plane-waves. A problem with the aforementioned discretization scheme is that the packet’s total angular momentum does {\it not} appear to converge toward the well-known continuum limit when $L\rightarrow\infty$\cite{Masud3}.

In accordance with the standard method of computing the angular momentum for a smooth (i.e., continuous and differentiable) EM wavepacket\cite{Cohen, Grynberg,Masud1}, those pairs of plane-waves that participate in generating the packet’s angular momentum are localized in the k-space, in the sense that a plane-wave and its immediate neighbors (in k-space) are the only pairs whose mutual interference contributes to the packet’s overall angular momentum\cite{Masud3}. Stated differently, the interference between two plane-waves whose k-vectors are not near-neighbors should give rise to an angular momentum density $\mathbf{j}(\mathbf{r},t)$ that averages out to zero when integrated over the entire xyz-space. In the case of spin angular momentum, one can show that the E-fields of adjacent plane-waves contribute in an “additive” way, so that each plane-wave member of the continuum appears to possess its own spin angular momentum—even though, in reality, this angular momentum arises from interference between each plane-wave and its adjacent k-space neighbors\cite{Masud3}. In contrast, in the case of orbital angular momentum, the E-fields of adjacent plane-waves are found to contribute in a “subtractive” way, so that the packet’s orbital angular momentum cannot be apportioned among the individual plane-wave members of the continuum; rather, it is the k-space gradients of the E-field components (i.e., local derivatives of $E_x, E_y, E_z$ with respect to $k_x, k_y, k_z)$ that participate in the creation of the packet’s orbital angular momentum\cite{Masud3}. This localization of the angular momentum within the k-space—be it the association of spin with individual k-vectors, or the allocation of orbital angular momentum in proportion to the k-space gradient of the EM field—is something that does not emerge naturally from a discretized k-space in the limit when $L\rightarrow\infty$.

In the present paper, we address the discrepancy between the discrete and continuum formulations of the EM angular momentum by examining a simplified model in which a scalar wave in two-dimensional (2D) Cartesian $xy$-space exhibits a similar discrepancy.

In Sec.2, we study the solution of the scalar wave equation in a homogeneous and isotropic 2D space, derive expressions for the energy $\mathcal{E}$, linear momentum $\mathcal{P}$, and angular momentum $\mathcal{J}$ of the corresponding wavepackets, and explore the conditions under which $\mathcal{E}$, $\mathcal{P}$, and $\mathcal{J}$ are conserved.

In Secs.3 and 4,  we expand the 2D field into plane-wave modes, and proceed to take a closer look at the conserved entities energy, linear momentum, and angular momentum, first in the infinite continuum of the $k_x k_y$-space, and next as the continuum limit, $L\rightarrow \infty$, of  an $L\cross L$ square cavity in the $xy$-plane harboring a spectrum of 2D plane-waves in a regularly discretized $k_x k_y$-space. 

 As compared to the electromagnetic field, the scalar field has, as expected, only orbital angular momentum. However, in spite of this important difference between the electromagnetic field and scalar field, it will be seen in Sec.5 that the continuum limit of the scalar 2D wavepacket’s angular momentum $\mathcal{J}$ in the discretized $k_x k_y$-space exhibits the same type of discrepancy as mentioned earlier in the context of 3D vector EM fields in the $xyz$-space.
 
 A conjecture that promises to resolve the discrepancy is proposed toward the end of Sec.5.

The proposed conjecture involves an operation performed on a function of a single variable, say, f(x), that combines a large number of samples taken from f(x) at regular intervals $\triangle=\frac{2\pi}{L}$, for instance, at $x_n=x_0\pm n\triangle$ for all integer values of $n$. The operator has the appearance of an infinite range difference operator.

The conjecture is proved in section 6, and reveal that the highly non-local operations in the discretized k-space that is needed to express the angular momentum $\mathcal{J}$ of a wavepacket (as discussed in Sec.5), do indeed yield the desired continuum result in the limit when $L\rightarrow\infty$, for all spectra for which the continuum formula for the total angular momentum make sense.

\section{Conserved quantities  for a 2D scalar field}
The defining equation for a 2D scalar field is the 2D linear wave equation
\begin{align}
\partial_{tt} u-c^2\nabla^2 u=0,\label{2.0.1}
\end{align}
where $u=u(t,x,y)$ is the scalar field and $\nabla^2=\partial_{xx}+\partial_{yy}$.

It is well known that the 2D wave equation is variational with Lagrangian density given by 
\begin{align}
\mathcal{L}=\frac{1}{2}u_t^2-\frac{1}{2}(\nabla u)^2.\label{2.0.2}
\end{align}

Observe that the Lagrangian density is of the general type
\begin{align*} 
\mathcal{L}=\mathcal{L}(t,x,y,u,u_t,u_x,u_y).
\end{align*}
In the context of Noether's theorem\cite{Weinberg,Per}, the Lagrangian
\begin{align}
    L[u]=\int_{\mathbb{R}^3} dt dx dy\; \mathcal{L}(t,x,y,u,u_t,u_x,u_y),\label{2.0.4}
\end{align}
corresponding to a Lagrangian density  of this type, is said to be ${\it invariant}$ under an infinitesimal variation of the form 
\begin{align*}
    u(t,x,y)\rightarrow u(t,x,y)+\epsilon \eta(t,x,y),
\end{align*}
if there exists functions $F_1, F_2$ and $F_3$ such that
\begin{align}
    \frac{\partial\mathcal{L}}{\partial u} \eta+ \frac{\partial\mathcal{L}}{\partial{u_t}} \eta_t+ \frac{\partial\mathcal{L}}{\partial{u_x}} \eta_x+ \frac{\partial\mathcal{L}}{\partial{u_y}} \eta_y=\partial_t F_1+\partial_x F_2+\partial_y F_3.\label{2.0.6}
\end{align}
Given this, there exists a local conservation law of the form
\begin{align*}
    \partial_t j_1+\partial_x j_2+\partial_y j_3=0,
\end{align*}
where the components of the Noether ${\it  current}$ are given by
\begin{align}
    j_1&=F_1-\eta\frac{\partial\mathcal{L}}{\partial{u_t}},\nonumber\\
    j_2&=F_2-\eta\frac{\partial\mathcal{L}}{\partial{u_x}},\nonumber\\
    j_3&=F_3-\eta\frac{\partial\mathcal{L}}{\partial{u_y}}.\label{2.0.8}
\end{align}
This is Noether's theorem for Lagrangians of the type (\ref{2.0.4}).

Consider now  an infinitesimal time translation of the form
\begin{align}
    t\rightarrow t+\epsilon.\label{2.1.1}
\end{align}
This infinitesimal time translation induces a variation on scalar fields of the form 
\begin{align}
    u(t,x,y)\rightarrow u(t,x,y)+\epsilon u_t(t,x,y).\label{2.1.2}
\end{align}
Thus, for this case 
\begin{align*}
\eta(t,x,y)=u_t(t,x,y).
\end{align*}
Using (\ref{2.0.2}), we now have 
\begin{align*}
     &\frac{\partial\mathcal{L}}{\partial u} \eta+ \frac{\partial\mathcal{L}}{\partial{u_t}} \eta_t+ \frac{\partial\mathcal{L}}{\partial{u_x}} \eta_x+ \frac{\partial\mathcal{L}}{\partial{u_y}} \eta_y\nonumber\\
     =&u_tu_{tt}-c^2u_xu_{tx}-c^2u_yu_{ty}\nonumber\\
     =&\partial_t(\frac{1}{2}u_t^2-\frac{1}{2}c^2u_x^2-\frac{1}{2}c^2u_y^2)\nonumber\\
     =&\partial_t\mathcal{L}.
\end{align*}
Thus, according to (\ref{2.0.6}), the Lagrangian (\ref{2.0.4}), corresponding to the Lagrangian density (\ref{2.0.2}) for the 2D wave equation (\ref{2.0.1}), is invariant under the variation (\ref{2.1.2}).  For this case we clearly have 
\begin{align*}
    F_1&=\mathcal{L},\nonumber\\
    F_2&=F_3=0.
\end{align*}
The components of the Noether current corresponding to the infinitesimal time translation (\ref{2.1.1}) are thus
\begin{align*}
    j_1=&F_1-\eta\frac{\partial\mathcal{L}}{\partial u_t}\nonumber\\
    =&(\frac{1}{2}u_t^2-\frac{1}{2}c^2u_x^2-\frac{1}{2}c^2u_y^2)-u_t^2\nonumber\\
    =&-(\frac{1}{2}u_t^2+\frac{1}{2}c^2u_x^2+\frac{1}{2}c^2u_y^2),\nonumber\\
    j_2=&F_2-\eta\frac{\partial\mathcal{L}}{\partial u_x}\nonumber\\
    =&(-u_t)(-c^2 u_x)=c^2 u_t u_x,\nonumber\\
    j_3=&F_3-\eta\frac{\partial\mathcal{L}}{\partial u_y}\nonumber\\
    =&(-u_t)(-c^2 u_y)=c^2 u_t u_y.
\end{align*}
Defining for convenience the quantities
\begin{align*}
    e(t,\mathbf{x})=&\frac{1}{2}u_t^2+\frac{1}{2}c^2u_x^2+\frac{1}{2}c^2u_y^2,\nonumber\\
    \mathbf{j}(t,\mathbf{x})=&-c^2u_t\nabla u,
\end{align*}
the local conservation law, corresponding to time translation, can be written in the form
\begin{align}
    \partial_t e+\nabla\cdot \mathbf{j}=0.\label{2.1.8}
\end{align}
The positive quantity $e(t,\mathbf{x})$ is by definition the energy density, and $j(t,\mathbf{x})$ the energy flux density for the scalar field.

For any domain $A$ in the plane with boundary curve $S$ and unit normal on $S$, pointing out of $A$, we get from (\ref{2.1.8}) that
\begin{align*}
    \frac{d\mathcal{E}}{dt}=-\int_S dl\; \mathbf{j}\cdot\mathbf{n}=c^2\int_S dl\; u_t\nabla u\cdot\mathbf{n}
\end{align*}
Here $\mathcal{E}$ is the, time dependent, total energy of the scalar field inside the domain $A$
\begin{align}
    \mathcal{E}(t)=\int_A d\mathbf{x}\; e(t,\mathbf{x}).\label{2.1.10}
\end{align}
Whether or not the energy of the scalar field inside $A$ is conserved, depends on the domain, and the boundary conditions the scalar field satisfies at the boundary of that domain.

The energy will clearly be conserved if  we assume that the domain D is so large, or the scalar field so localized, that it is negligible in a sufficiently small region around the boundary. 

If we cannot assume that the scalar field is negligible in a sufficiently small region around the boundary, the energy can still be conserved if, on any section of the boundary, either $u$ is constant in time, or $\nabla u\cdot\mathbf{n}=0$.

In Appendix A, using the same approach, we find the conserved quantities corresponding to  infinitesimal space translations and rotations. These are, by definition,  total linear momentum
\begin{align}
\mathcal{P} = -\int_A d\mathbf{x}\; u_t \; \grad{u},\label{2.2.10} 
\end{align}
and total  angular momentum
\begin{align}
\mathcal{J} =  -\int_A d\mathbf{x}\; u_t\; \mathbf{x}\cross\nabla u.\label{2.3.11} 
\end{align}
In Appendix A we also show how the linear momentum flux and angular momentum flux for the scalar field follow directly from the expression for the Noether current. These formulas are the analogues to the linear and angular momentum fluxes\cite{Barnett1} for the electromagnetic field.

\section{The continuum limit for energy}
In this section we will find the formulas for the total energy for a scalar field in the whole plane, which we will here denote by the {\it continuum},  and for a square cavity of side length $L$, in terms of continuous and discrete modes, respectively.  We will then investigate how the formula for the total energy in the cavity approaches the one for the continuum in the continuum limit, defined by letting $L$ approach infinity. Detailed calculations and derivations from this section are left out of the main text and can be found in Appendix B.

\subsection{The continuum}
For the continuum we will look for solutions to the equation for the scalar field
\begin{align}
\partial_{tt} u-c^2\nabla^2 u=0,\label{3.1.1}
\end{align}
of the form
\begin{align*}
    u(t,\mathbf{x})=e^{i(\mathbf{k}\cdot\mathbf{x}-\omega t)}.
\end{align*}
Here $\mathbf{k}=(k_x,k_y)$, and the frequency $\omega$ is positive.  For modes to exist we must have 
\begin{align*}
    \omega^2=c^2 k_x^2+c^2 k_y^2.
\end{align*}
Given that $\omega>0$, this equation has exactly two solutions
\begin{align*}
    \omega=\pm \omega(k),
\end{align*}
where
\begin{align*}
   \omega(k)&=ck,\nonumber\\
   k&\equiv\norm{\mathbf{k}}=(k_x^2+k_y^2)^{\frac{1}{2}}.
\end{align*}
These are the two branches of the dispersion relation for the scalar field equation (\ref{3.1.1}).

The two different branches of the dispersion relation give rise to two different families of modes
\begin{align*}
    e^{i(\mathbf{k}\cdot\mathbf{x}-\omega(k) t)},\;\; e^{i(\mathbf{k}\cdot\mathbf{x}+\omega(k) t)}.
\end{align*}
These modes, also known as {\it plane waves},  are known to be complete, and using them, we can write the general complex solution to equation (\ref{3.1.1}) in the form
\begin{align}
    u(t,\mathbf{x})=\int d\mathbf{k}\;e^{i\mathbf{k}\cdot\mathbf{x}}\left\{A(\mathbf{k})e^{-i\omega(k) t}+B(\mathbf{k})e^{i\omega(k) t}\right\},\label{3.1.7}
\end{align}
where  $A(\mathbf{k})$ and $B(\mathbf{k})$ are the {\it spectral amplitudes} defining the solution.  In this formula the integral goes over the whole 2D space of $\mathbf{k}$-vectors.

We are here focused on real solutions, and the reality of $u(t,\mathbf{x})$ is ensured by imposing the condition 
\begin{align}
    B(\mathbf{k})=A(-\mathbf{k})^*,\label{3.1.8}
\end{align}
on the spectral amplitudes.

Let us now specialize the general space-time formula (\ref{2.1.10}) for the total energy of the scalar field  in a domain $A$, to the case of the continuum, $A=\mathbb{R}^2$, 
\begin{align}
    \mathcal{E}(t)=\int_{\mathbb{R}^2} d\mathbf{x}\; e(t,\mathbf{x})=\int_{\mathbb{R}^2} d\mathbf{x}\; \left(\frac{1}{2}u_t^2+\frac{1}{2}c^2u_x^2+\frac{1}{2}c^2u_y^2\right).\label{2.1.10.1}
\end{align}
From the mode expansion (\ref{3.1.7}) of the scalar field we immediately  have that
\begin{align}
    u_t=&\int d\mathbf{k}\;(-i\omega(k))e^{i\mathbf{k}\cdot\mathbf{x}}\left\{A(\mathbf{k})e^{-i\omega(k) t}-B(\mathbf{k})e^{i\omega(k) t}\right\},\nonumber\\
     \nabla u=&\int d\mathbf{k}\;(i\mathbf{k})e^{i\mathbf{k}\cdot\mathbf{x}}\left\{A(\mathbf{k})e^{-i\omega(k) t}+B(\mathbf{k})e^{i\omega(k) t}\right\},\label{3.1.1.3}
\end{align}
and, inserting these expansions into (\ref{2.1.10.1}), it is straightforward to find that 

\begin{align*}
    \mathcal{E}&=(2\pi)^2\int d\mathbf{k}\;\omega^2\left\{\abs{A(\mathbf{k})}^2+ \abs{B(\mathbf{k})}^2 \right\}.
\end{align*}

Note that the formula for total energy of the scalar field, expressed in terms of spectral amplitudes,  is time invariant, as it should be.  Also note that the total energy in the scalar field is simply a sum over the individual energies in all the plane waves that are used to construct the field. This is what our intuition leads us to expect as well.

\subsection{A square cavity}
For a scalar field in a square cavity
\begin{align*}
    A=[-\frac{L}{2},\frac{L}{2}]\cross[-\frac{L}{2},\frac{L}{2}],
\end{align*}
the modes and dispersion relation are the same as for the continuum
\begin{align*}
    u(t,\mathbf{x})=e^{i(\mathbf{k}\cdot\mathbf{x}-\omega(k) t)},\;\; e^{i(\mathbf{k}\cdot\mathbf{x}+\omega(k) t)},
\end{align*}
where $\omega(k)=ck$.

However, not all these plane waves satisfy the required periodic boundary conditions
\begin{align*}
    u(t,x-\frac{L}{2},y)&= u(t,x+\frac{L}{2},y),\nonumber\\
     u(t,x,y-\frac{L}{2})&= u(t,x,y+\frac{L}{2}).
\end{align*}
These conditions lead to a discrete set of possible $\mathbf{k}$-vectors, and corresponding frequencies
\begin{align}
    \mathbf{k}_{\mathbf{n}}&=\left(\frac{2\pi}{L}\right)(n_x,n_y),\nonumber\\
    \omega_{n}&=\left(\frac{2\pi c}{L}\right)(n_x^2+n_y^2)^{\frac{1}{2}}.\label{3.2.4}
\end{align}
Here the quantities $n_x$ and $n_y$ run over all integers, and we have defined $n=\norm{\mathbf{n}}=(n_x^2+n_y^2)^{\frac{1}{2}}$.

The plane wave modes for the scalar field in the square cavity are then of the form
\begin{align*}
    a_{\mathbf{n}}e^{i( \mathbf{k}_{\mathbf{n}}\cdot\mathbf{x}-\omega_{n}t)},\;\;\;b_{\mathbf{n}}e^{i( \mathbf{k}_{\mathbf{n}}\cdot\mathbf{x}+\omega_{n}t)}.
\end{align*}
Here, $a_{\mathbf{n}}$ and $b_{\mathbf{n}}$ are the discrete spectral amplitudes.

The general real solution to the scalar field equation for a square cavity is then of the form
\begin{align}
    u(t,\mathbf{x})&=\sum_{\mathbf{n}}e^{i \mathbf{k}_{\mathbf{n}}\cdot\mathbf{x}}\left\{ a_{\mathbf{n}}e^{-i\omega_{\mathbf{n}}t}+  b_{\mathbf{n}}e^{i\omega_{\mathbf{n}}t}\right\},\label{3.2.6}
\end{align}
where we have imposed the reality conditions
\begin{align*}
    b_{\mathbf{n}}=a_{-\mathbf{n}}^*.
\end{align*}
We now want to find the formula for the conserved energy in terms of the discrete modes for the scalar field.
In this context it is useful to know that the total energy of the scalar field in the cavity is conserved. We do not need to assume that the scalar field vanishes in a region close to the boundary. This occurs because the energy  flux density satisfy periodic boundary conditions, if the scalar field does. The same is true for linear momentum, but it is not true for angular momentum. The culprit for angular momentum is the factor $\mathbf{x}^{\perp}$ occurring in the angular momentum flux density. It destroys periodicity.
Thus in order to ensure that all three of total energy, linear momentum, and angular momentum, are conserved, we  must assume that the scalar field decays so fast away from the origin that it is vanishingly  small close to the boundary of the cavity. This is what we will assume from now on. The expressions for linear momentum- and angular momentum flux densities can be found in Appendix A.

The total energy of the scalar field inside the square cavity is given by the formula
\begin{align}
    \mathcal{E}_L(t)=\int_{A} d\mathbf{x}\; \left(\frac{1}{2}u_t^2+\frac{1}{2}c^2u_x^2+\frac{1}{2}c^2u_y^2\right).\label{2.1.10.2}
\end{align}
From the expansion of the scalar field in terms of discrete cavity modes (\ref{3.2.6}) we have
\begin{align}
    u_t=\sum_{\mathbf{n}}\;(-i\omega_{n})e^{i \mathbf{k}_{\mathbf{n}}\cdot\mathbf{x}}\left\{ a_{\mathbf{n}}e^{-i\omega_{n}t}-  b_{\mathbf{n}}e^{i\omega_{n}t}\right\},\nonumber\\
     \nabla u=\sum_{\mathbf{n}}\;(i\mathbf{k}_{\mathbf{n}})e^{i \mathbf{k}_{\mathbf{n}}\cdot\mathbf{x}}\left\{ a_{\mathbf{n}}e^{-i\omega_{n}t}+  b_{\mathbf{n}}e^{i\omega_{n}t}\right\}.\label{3.2.1.3}
\end{align}
Inserting these formulas into the expression for the total energy in the square cavity (\ref{2.1.10.2}), we find, using derivations similar those in the continuum case, that the total energy in the cavity is given by 
    \begin{align}
         \mathcal{E}_L=L^2\sum_{\mathbf{n}}\;\omega_{n}^2\left\{\abs{a_{\mathbf{n}}}^2+\abs{b_{\mathbf{n}}}^2\right\}.\label{3.2.1.8}
    \end{align}
As in  the continuum case, we note that the total energy of the scalar field in the cavity is a sum over the energies residing in individual plane wave modes that the scalar field.

\subsection{Continuum limit}

    For a fixed size of the  cavity, the allowed $\mathbf{k}$-vectors form a uniform square grid in the 2D space of all possible vectors. This square grid is unbounded and the density of the grid is fixed by the distance between nearest neighbors. This distance will be denoted by $\triangle$ 
    \begin{align*}
        \triangle=\left(\frac{2\pi}{L}\right).
    \end{align*}
    Thus we have 
    \begin{align}
        \triangle k_x=\triangle k_y=\triangle.\label{4.1.3}
    \end{align}
    Note that in the continuum limit, which is defined by letting $L$ approach infinity, the distance $\triangle$ approaches zero. 
    
    Since the spectral amplitudes, $A(\mathbf{k})$ and $B(\mathbf{k})$,  determining the scalar field for the continuum case (\ref{3.1.7}), are ${\it densities}$ in $\mathbf{k}$-space, we must make the identification 
   \begin{align}
     a_{\mathbf{n}}&=\triangle \mathbf{k}\;A(\mathbf{k}_{\mathbf{n}}),\nonumber\\
      b_{\mathbf{n}}&=\triangle \mathbf{k}\;B(\mathbf{k}_{\mathbf{n}}),\label{4.1.4}
   \end{align}
   where
   \begin{align}
       \triangle \mathbf{k}=\triangle k_x\triangle k_y=\triangle^2.\label{4.1.5}
   \end{align}
   Inserting (\ref{4.1.4}) into (\ref{3.2.1.8}), we get
   \begin{align*}
         \mathcal{E}_L=&L^2\triangle^2\sum_{\mathbf{n}}\;\triangle\mathbf{k}\;\omega_{n}^2\left\{\abs{A(\mathbf{k}_{\mathbf{n}})}^2+\abs{B(\mathbf{k}_{\mathbf{n}})}^2\right\}\nonumber\\
        &=(2\pi)^2\sum_{\mathbf{n}}\;\triangle\mathbf{k}\;\omega_{n}^2\left\{\abs{A(\mathbf{k}_{\mathbf{n}})}^2+\abs{B(\mathbf{k}_{\mathbf{n}})}^2\right\}\nonumber\\
        &\rightarrow (2\pi)^2\int d\mathbf{k}\;\omega^2\left\{\abs{A(\mathbf{k})}^2+ \abs{B(\mathbf{k})}^2 \right\}\nonumber\\
        &=\mathcal{E},
   \end{align*}
   in the continuum limit, $L\rightarrow\infty$.

   Thus,  in the continuum limit, the discrete mode formula for total energy of the scalar field in the square cavity approaches the continuum mode formula. This is exactly what one would expect.

\section{The continuum limit for linear momentum}
In this section we derive the formula for the linear momentum in terms of continuum modes, and for a square cavity in terms of discrete modes, and show that the first is the continuum limit of the second.  Detailed calculations and derivations in this section are left out of the main text and can be found in Appendix C.

From expressions (\ref{3.1.1.3}), and the expression for the total linear momentum of the scalar field (\ref{2.2.10}), using calculations similar to those for the energy, we find that the formula for total linear momentum  of the scalar field in the continuum, expressed in terms of spectral amplitudes,  is time invariant, as it should be, and takes the form
\begin{align*}
    \mathcal{P}&= (2\pi)^2\int d\mathbf{k}\;\omega\left\{\mathbf{k}\abs{A(\mathbf{k})}^2-\mathbf{k}\abs{B(\mathbf{k})}^2 \right\},
\end{align*}
and that the total linear momentum of the scalar field in a square cavity, expressed in terms of discrete cavity modes takes the form 
\begin{align*}
    \mathcal{P}_L= L^2\sum_{\mathbf{n}}\;\omega_{n}\left\{\mathbf{k}_{\mathbf{n}}\abs{a_{\mathbf{n}}}^2-\mathbf{k}_{\mathbf{n}}\abs{b_{\mathbf{n}}}^2\right\}.
\end{align*}
As was the case for the total energy, we  note that the total linear momentum content of the scalar field is simply a sum over the individual momenta of all the plane waves that are used to construct the field. This is again what intuition leads us to expect.

Proceeding as in for the case of the total energy we now have
\begin{align*}
     \mathcal{P}_L&= L^2\triangle^2\sum_{\mathbf{n}}\;\triangle\mathbf{k}\;\omega_{n}\left\{\mathbf{k}_{\mathbf{n}}\abs{A(\mathbf{k}_{\mathbf{n}})}^2-\mathbf{k}_{\mathbf{n}}\abs{B(\mathbf{k}_{\mathbf{n}})}^2\right\}\nonumber\\
     &=(2\pi)^2\sum_{\mathbf{n}}\;\triangle\mathbf{k}\;\omega_{n}\left\{\mathbf{k}_{\mathbf{n}}\abs{A(\mathbf{k}_{\mathbf{n}})}^2-\mathbf{k}_{\mathbf{n}}\abs{B(\mathbf{k}_{\mathbf{n}})}^2\right\}\nonumber\\
     &\rightarrow(2\pi)^2\int d\mathbf{k}\;\omega\left\{\mathbf{k}\abs{A(\mathbf{k})}^2-\mathbf{k}\abs{B(\mathbf{k})}^2 \right\}\nonumber\\
     &=\mathcal{P}.
\end{align*}
Thus,  in the continuum limit, the formula for the total momentum of the scalar field inside the cavity approaches the continuum mode formula. This is also just what we should expect.

\section{The continuum limit for angular momentum}
In this section we will finally investigate the continuum limit for the angular momentum in a square cavity.
We have  previously made a note of  the fact that energy and momentum in the scalar field reside in each individual plane wave comprising  the field. This is the reason why  deriving the continuum limit for the energy and linear momentum was so unproblematic. For the angular momentum, which, as we shall see,  does {\it not} reside in the individual plane wave modes, the continuum limit is much harder to determine.
 Detailed calculations and derivations for this section are left out of the main text and can be found in Appendix D.

\subsection{The continuum}
Using the formulas  (\ref{3.1.1.3}), and the expression for the total angular momentum of the scalar field (\ref{2.3.11}), we find using calculations similar to the ones for energy and linear momentum, that  
he formula for total angular momentum  of the scalar field in the continuum, expressed in terms of spectral amplitudes, takes the form
\begin{align*}
    \mathcal{J}&=-i(2\pi)^2\int d\mathbf{k}\;\omega\left\{A(\mathbf{k})^*(\mathbf{k}\cross\nabla_{\mathbf{k}})A(\mathbf{k})-B(\mathbf{k})^*(\mathbf{k}\cross\nabla_{\mathbf{k}})B(\mathbf{k})\right\}.
\end{align*}
It is perhaps not immediately clear, as it was for the formulas for the total energy and linear momentum, that the formula for the total angular momentum actually produces a real quantity.

Using the properties (\ref{3.1.8}) for the spectral amplitudes $A$ and $B$, we have however,
\begin{align*}
    \mathcal{J}^*&=i(2\pi)^2\int d\mathbf{k}\;\omega\left\{A(\mathbf{k})(\mathbf{k}\cross\nabla_{\mathbf{k}})A(\mathbf{k})^*-B(\mathbf{k})(\mathbf{k}\cross\nabla_{\mathbf{k}})B(\mathbf{k})^*\right\}\nonumber\\
   &=i(2\pi)^2\int d\mathbf{k}\;\omega\left\{A(-\mathbf{k})(\mathbf{k}\cross\nabla_{\mathbf{k}})A(-\mathbf{k})^*-B(-\mathbf{k})(\mathbf{k}\cross\nabla_{\mathbf{k}})B(-\mathbf{k})^*\right\}\nonumber\\ 
    &=-i(2\pi)^2\int d\mathbf{k}\;\omega\left\{A(\mathbf{k})^*(\mathbf{k}\cross\nabla_{\mathbf{k}})A(\mathbf{k})-B(\mathbf{k})^*(\mathbf{k}\cross\nabla_{\mathbf{k}})B(\mathbf{k})\right\}\nonumber\\ 
     &= \mathcal{J},
\end{align*}
so $ \mathcal{J}$ is indeed real.
Note that, contrary to the cases of energy and linear momentum, the total angular momentum of a scalar field is not simply the sum of the angular momenta of the individual modes, but rather contains a gradient in $\mathbf{k}$-space. Thus, individual modes do not have angular momentum. Angular momentum rather arise from local relations {\it between} individual plane wave modes.

\subsection{A square cavity}
In this section we will find a formula for the total angular momentum of the scalar field inside the square cavity. 
Using expressions (\ref{3.2.1.3}) we have
\begin{align}
    \mathcal{J}_L&=-\int_A d\mathbf{x}\;u_t\;\mathbf{x}\cross\nabla u\nonumber\\
    &=-\sum_{\mathbf{n}\mathbf{n}'}\;\omega'S_{\mathbf{n}'\mathbf{n}}\nonumber\\
&\left\{a'ae^{-i(\omega+\omega')t}+a'be^{-i(\omega-\omega')t}-b'ae^{i(\omega-\omega')t}-b'be^{i(\omega+\omega')t} \right\},\label{3.2.3.1}
\end{align}
 where we are using the compact notation $a=a_{\mathbf{n}},a'=a_{\mathbf{n}'}$, etc, and 
where the symbol $S_{\mathbf{n}'\mathbf{n}}$ is defined by
\begin{align*}
  S_{\mathbf{n}'\mathbf{n}}= \int_A d\mathbf{x}\;( \mathbf{x}\cross\mathbf{k})e^{i (\mathbf{k}_{\mathbf{n}}+\mathbf{k}_{\mathbf{n}'})}.
\end{align*}
Evaluating  this integral is an elementary but somewhat tedious exercise, and we will not detail the solution in this report. The final expression is rather simple and given by
\begin{align}
  S_{\mathbf{n}'\mathbf{n}}= iL^2\left(n_x\delta_{n_x+n_x'}\frac{(-1)^{\abs{n_y+n_y'}}}{n_y+n_y'}-n_y\delta_{n_y+n_y'}\frac{(-1)^{\abs{n_x+n_x'}}}{n_x+n_x'}\right).
  \label{3.2.3.3}
\end{align}
Note that in order for this formula to be correct, we must let $\frac{(-1)^0}{0}$ be assigned the value $0$, whenever it occurs.

Formula (\ref{3.2.3.1}) for the total angular momentum consists of four terms. Let us denote the contribution to the angular momentum from these four terms by $ \mathcal{J}_i, i=1,\dots 4$.

We now insert the expression (\ref{3.2.3.3}) into the first of these terms, $\mathcal{J}_1$, and rewrite it into the following  particular form that we will need for doing the continuum limit in the next section
\begin{align}
     \mathcal{J}_1
    &=i L^2\sum_{\mathbf{n}}a_{-\mathbf{n}}e^{-i\omega_{n}t}\left\{ n_x \Lambda_y(f_1)_{\mathbf{n}}-n_y\Lambda_x(f_1)_{\mathbf{n}}\right\}.\label{3.2.3.10}
\end{align}
Here $f_1$ is a sequence given by 
\begin{align*}
    (f_1)_{\mathbf{n}}=\omega_{n}a_{\mathbf{n}}e^{-i\omega_{n}t}.
\end{align*}

In the expression for $ \mathcal{J}_1$,  $\Lambda_x$, and $\Lambda_y$ are operators acting on doubly infinite sequences 
$\alpha_{\mathbf{n}}=\alpha_{n_x,n_y}$ according to 
\begin{align*}
    \Lambda_x(\alpha)_{nx,ny}&=\sum_s\;P_s \alpha_{n_x+s,n_y},\nonumber\\
    \Lambda_y(\alpha)_{n_x,n_y}&=\sum_s\;P_s \alpha_{n_x,n_y+s},
\end{align*}
where the symbol $P_q$ is defined by 
\begin{align*}
    P_q&=\frac{(-1)^{\abs{q}}}{q},\;\;\;q\neq 0,\nonumber\\
    P_0&=0.
\end{align*}

In a similar way we get the following formulas for the other components of the total angular momentum, $ \mathcal{J}_j,j=2,3,4$
\begin{align*}
     \mathcal{J}_2&=i L^2\sum_{\mathbf{n}}b_{-\mathbf{n}}e^{-i\omega_{n}t}\left\{ n_x \Lambda_y(f_2)_{\mathbf{n}}-n_y\Lambda_x(f_2)_{\mathbf{n}}\right\},\nonumber\\
     \mathcal{J}_3&=-i L^2\sum_{\mathbf{n}}a_{-\mathbf{n}}e^{i\omega_{n}t}\left\{ n_x \Lambda_y(f_3)_{\mathbf{n}}-n_y\Lambda_x(f_3)_{\mathbf{n}}\right\},\nonumber\\
     \mathcal{J}_4&=-i L^2\sum_{\mathbf{n}}b_{-\mathbf{n}}e^{i\omega_{n}t}\left\{ n_x \Lambda_y(f_4)_{\mathbf{n}}-n_y\Lambda_x(f_4)_{\mathbf{n}}\right\},
\end{align*}
where
\begin{align*}
    (f_2)_{\mathbf{n}}&=\omega_{\mathbf{n}}a_{\mathbf{n}}e^{i\omega_{n}t},\nonumber\\
    (f_3)_{\mathbf{n}}&=\omega_{\mathbf{n}}b_{\mathbf{n}}e^{-i\omega_{n}t},\nonumber\\
    (f_4)_{\mathbf{n}}&=\omega_{n}b_{\mathbf{n}}e^{i\omega_{n}t}.
\end{align*}
The total angular momentum of the scalar field in the cavity is now given by 

\begin{align*}
     \mathcal{J}_L=\sum_{j=1}^4\; \mathcal{J}_j.
\end{align*}
From this formula, it is clear that in parallel with the continuum case, the angular momentum in the cavity resides not in the individual plane waves, but rather in the relations between the waves. In contrast to what we observed for the continuum case, here, the angular momentum does not arise from the relations between nearby modes. The formula seems to be telling us that the total angular momentum is the sum over quantities, these are the $\Lambda$ operators, that quantify relations between each individual plane wave and {\it all} other plane waves. For a finite sized cavity, no matter the size, 
angular momentum seems to be totally delocalized in $\mathbf{k}$-space. 

\subsection{The continuum limit}\label{AngularMomentumContinuumLimit}

Recall that the formula for the total angular momentum for the scalar field in a square cavity is the sum of four terms, $\mathcal{J}_i,i=1,\dots,4$. 

The formula for $\mathcal{J}_1$ has been found to be
\begin{align}
    \mathcal{J}_1&=i L^2\sum_{\mathbf{n}}a_{-\mathbf{n}}e^{-i\omega_{n}t}\left\{ n_x \Lambda_y(f_1)_{\mathbf{n}}-n_y\Lambda_x(f_1)_{\mathbf{n}}\right\}.\label{4.3.0.1}
\end{align}
In this section it is convenient to rewrite the operators $\Lambda_x$ and $\Lambda_y$. Using the fact the $P_{-s}=-P_s$, we have
\begin{align*}
     \Lambda_x(a)_{\mathbf{n}}&=\sum_s\;P_s a_{n_x+s,n_y}\nonumber\\
     &=\sum_{s=-\infty}^{-1}\;P_s a_{n_x+s,n_y}+\sum_{s=1}^{\infty}\;P_s a_{n_x+s,n_y}\nonumber\\
      &=\sum_{s=1}^{\infty}\;P_s \left(a_{n_x+s,n_y}-a_{n_x-s,n_y}\right).
\end{align*}
The operator $\Lambda_y$ is rewritten in a similar way.

Now, using formulas (\ref{4.1.3}), (\ref{4.1.4}) and (\ref{4.1.5}), we have
\begin{align}
    (f_1)_{\mathbf{n}}&=\omega_{n}a_{\mathbf{n}}e^{-i\omega_{n}}\nonumber\\
    &=\triangle\mathbf{k}\;\omega(k_n)A(\mathbf{k}_{\mathbf{n}})e^{-i\omega(k_n) t}.\label{4.3.0.3}
\end{align}
We now define a function $F_1$ in $\mathbf{k}$-space by the formula
\begin{align}
    F_1(\mathbf{k})=\omega(k)A(\mathbf{k})e^{-i\omega(k) t}.\label{4.0.4}
\end{align}
Given this, (\ref{4.3.0.3}) can be written in the form
\begin{align*}
     (f_1)_{\mathbf{n}}&=\triangle\mathbf{k}\;F_1(\mathbf{k}_{\mathbf{n}})\nonumber\\
     &\Updownarrow\nonumber\\
     (f_1)_{n_x,n_y}&=\triangle\mathbf{k}\;F_1(\triangle n_x,\triangle n_y) .
\end{align*}
We now prepare the operators $\Lambda_x$ and $\Lambda_y$ for passing to of the continuum limit.
\begin{align}
    (\Lambda_x)(f_1)_{\mathbf{n}}&=\sum_{s=1}^{\infty}\frac{(-1)^s}{s}\left ((f_1)_{n_x+s,n_y}-(f_1)_{n_x-s,n_y}\right)\nonumber\\
    &=\triangle\mathbf{k}\;\sum_{s=1}^{\infty}\frac{(-1)^s}{s}\left (F_1(\triangle(n_x+s),\triangle n_y)-F_1(\triangle(n_x-s),\triangle n_y)\right)\nonumber\\
     &=\triangle\mathbf{k}\;\sum_{s=1}^{\infty}\frac{(-1)^s}{s}\left (F_1(\triangle n_x+\triangle s,\triangle n_y)-F_1(\triangle n_x-\triangle s,\triangle n_y)\right)\nonumber\\
    &=-\triangle^3\;\mathbb{D}_x^L(F_1)(\mathbf{k}_{\mathbf{n}}),\label{4.3.0.5}
\end{align}
where we have defined a new operator $\mathbb{D}_x^L$, acting on functions $G(\mathbf{k})$ defined on $\mathbf{k}$-space  according to
\begin{align*}
   \mathbb{D}_x^L(G)(\mathbf{k})= \sum_{s=1}^{\infty}(-1)^{(s-1)}\frac{G(\mathbf{k}+(\triangle s,0))-G(\mathbf{k}-(\triangle s,0))}{\triangle s}.
\end{align*}
Treating the operator $\Lambda_y$ in the same way we find that 
\begin{align}
     (\Lambda_y)(f_1)_{\mathbf{n}}&=-\triangle^3\;\mathbb{D}_y^L(F_1)(\mathbf{k}_{\mathbf{n}}),\label{4.3.0.7}
\end{align}
where the operator $\mathbb{D}_y^L$ is defined by 
\begin{align*}
   \mathbb{D}_y^L(G)(\mathbf{k})= \sum_{s=1}^{\infty}(-1)^{(s-1)}\frac{G(\mathbf{k}+(0,\triangle s))-G(\mathbf{k}-(0,\triangle s))}{\triangle s}.
\end{align*}
\label{ExactContinuumLimitM1}
Inserting (\ref{4.3.0.5}) and (\ref{4.3.0.7}) into (\ref{4.3.0.1}), and again using (\ref{4.1.4}) and (\ref{4.1.5}) we have
\begin{align}
    &\mathcal{J}_1=i L^2\sum_{\mathbf{n}}a_{-\mathbf{n}}e^{-i\omega_{n}t}\left\{ n_x \Lambda_y(f_1)_{\mathbf{n}}-n_y\Lambda_x(f_1)_{\mathbf{n}}\right\}\nonumber\\
    &=-i (2\pi)^2\sum_{\mathbf{n}}\triangle\mathbf{k}\;A(-\mathbf{k}_{\mathbf{n}})e^{-i\omega(k_n)t}\left\{ \triangle n_x \mathbb{D}_y^L(F_1)(\mathbf{k}_{\mathbf{n}})-\triangle n_y \mathbb{D}_x^L(F_1)(\mathbf{k}_{\mathbf{n}})\right\}\nonumber\\
    &\rightarrow -i(2\pi)^2\int d\mathbf{k}A(-\mathbf{k})e^{-i\omega(k)t}\left\{ k_x \mathbb{D}_y(F_1)(\mathbf{k})- k_y \mathbb{D}_x(F_1)(\mathbf{k})\right\}.\label{4.3.0.9}
\end{align}
The two operators $\mathbb{D}_x$ and $\mathbb{D}_y$ are the continuum limits of  $\mathbb{D}_x^L$ and $\mathbb{D}_y^L$  in the sense that
\begin{align*}
    \mathbb{D}_x(G)(\mathbf{k})&=\lim_{L\rightarrow\infty} \left(\mathbb{D}_x^L(G)(\mathbf{k})\right)\nonumber\\
    &=\lim_{L\rightarrow\infty}\sum_{s=1}^{\infty}(-1)^{(s-1)}\frac{G(\mathbf{k}+(\triangle s,0))-G(\mathbf{k}-(\triangle s,0))}{\triangle s}, \nonumber\\
     \mathbb{D}_y(G)(\mathbf{k})&=\lim_{L\rightarrow\infty} \left(\mathbb{D}_y^L(G)(\mathbf{k})\right)\nonumber\\
     &=\lim_{L\rightarrow\infty}\sum_{s=1}^{\infty}(-1)^{(s-1)}\frac{G(\mathbf{k}+(0,\triangle s))-G(\mathbf{k}-(0,\triangle s))}{\triangle s}.
\end{align*}
It is clear from these formulas that the operators $\mathbb{D}_x^L(G)$ and  $\mathbb{D}_y^L(G)$ are highly non-local. For example, computing the value of $\mathbb{D}_x^L(G)(\mathbf{k})$, at some chosen point $\mathbf{k}$, will involve wave numbers arbitrarily distant from $\mathbf{k}$, no matter the value of $L$. It is not evident at all what the continuum limit of such highly non-local entities should be.

Nevertheless, in order to find the continuum limit for the total angular momentum in a square cavity, the precise nature of the two operators $ \mathbb{D}_x(G)$ and  $ \mathbb{D}_y(G)$ must be understood. 

\subsubsection{A simplifying assumption; the factor-of-two problem}
We start by observing that if we define the following operator, $\mathbb{D}$, acting on functions, $f(x)$, on the real line
\begin{align*}
    \mathbb{D}(f)(x)=\lim_{L\rightarrow\infty}\sum_{s=1}^{\infty}(-1)^{(s-1)}\frac{f(x+\triangle s)-f(x-\triangle s)}{\triangle s},
\end{align*}
then both $\mathbb{D}_x$ and $\mathbb{D}_y$ will be fully understood, if $\mathbb{D}$ is. Recall here that $\triangle=\frac{2\pi}{L}$.

We evidently can rewrite the previous formula into the form
\begin{align}
    \mathbb{D}(f)(x)=\lim_{L\rightarrow\infty}\frac{f(x+\triangle )-f(x-\triangle )}{\triangle }+\lim_{L\rightarrow\infty}R(f)(x,L),\label{4.3.1.2}
\end{align}
where we have defined 
\begin{align*}
    R(f)(x,L)=\sum_{s=2}^{\infty}(-1)^{(s-1)}\frac{f(x+\triangle s)-f(x-\triangle s)}{\triangle s}.
\end{align*}
 Observe that ${\it if}$  we postulate that the last term in (\ref{4.3.1.2}) vanishes
 \begin{align*}
     \lim_{L\rightarrow\infty}R(f)(x,L)=0,
 \end{align*}
 we get simply
\begin{align*}
   \mathbb{D}(f)(x)=2f'(x).
\end{align*}
Returning to the discussion in the previous section we conclude that, assuming the postulate, we have
\begin{align*}
    \mathbb{D}_x(G)(\mathbf{k})&=2\partial_{k_x}G(\mathbf{k}),\nonumber\\
     \mathbb{D}_y(G)(\mathbf{k})&=2\partial_{k_y}G(\mathbf{k}).
\end{align*}
Given this we have 
\begin{align*}
    \left\{ k_x \mathbb{D}_y(F_1)(\mathbf{k})- k_y \mathbb{D}_x(F_1)(\mathbf{k})\right\}&=2 \left\{ k_x \partial_{k_y}F_1(\mathbf{k})- k_y \partial_{k_x}F_1(\mathbf{k})\right\}\nonumber\\
    &=2(\mathbf{k}\cross\nabla_{\mathbf{k}})F_1(\mathbf{k}).
\end{align*}
Note that here, as elsewhere in this paper, we identify vectors pointing along the z-axis with their z-component.

Using the definition of the function $F_1$ from (\ref{4.0.4}) we have
\begin{align*}
   (\mathbf{k}\cross\nabla_{\mathbf{k}})F_1(\mathbf{k})&= (\mathbf{k}\cross\nabla_{\mathbf{k}})(\omega(k)A(\mathbf{k})e^{-i\omega(k) t})\nonumber\\
   &=\omega(k)\;(\mathbf{k}\cross\nabla_{\mathbf{k}})A(\mathbf{k})\;e^{-i\omega(k) t}.
\end{align*}
This formula is true because
\begin{align*}
    (\mathbf{k}\cross\nabla_{\mathbf{k}})\omega(k)=\frac{\omega'(k)}{\norm{\mathbf{k}}}(\mathbf{k}\cross\mathbf{k})=0.
\end{align*}
Thus, from our postulate, the continuum limit of  $ \mathcal{J}_1$ derived in  the previous section (\ref{4.3.0.7}), turns into
\begin{align*}
    \mathcal{J}_1\rightarrow -2i(2\pi)^2\int d\mathbf{k}\;\omega A(-\mathbf{k})(\mathbf{k}\cross\nabla_{\mathbf{k}})A(\mathbf{k})^{-2i\omega t}.
\end{align*}
In Appendix D, on page \pageref{VanishingIntegral}, we prove that an expression like this vanishes.
In a similar way we find that $\mathcal{J}_4$ vanishes in the continuum limit.

For $\mathcal{J}_2$, using the same approach as for $\mathcal{J}_1$,  we have 
\begin{align*}
    &\mathcal{J}_2=i L^2\sum_{\mathbf{n}}b_{-\mathbf{n}}e^{-i\omega_{n}t}\left\{ n_x \Lambda_y(f_2)_{\mathbf{n}}-n_y\Lambda_x(f_2)_{\mathbf{n}}\right\}\nonumber\\
    &=-i (2\pi)^2\sum_{\mathbf{n}}\triangle\mathbf{k}\;B(-\mathbf{k}_{\mathbf{n}})e^{-i\omega(k_n)t}\left\{ \triangle n_x \mathbb{D}_y^L(F_2)(\mathbf{k}_{\mathbf{n}})-\triangle n_y \mathbb{D}_x^L(F_2)(\mathbf{k}_{\mathbf{n}})\right\}\nonumber\\
    &\rightarrow -i(2\pi)^2\int d\mathbf{k} B(-\mathbf{k})e^{-i\omega(k)t}\left\{ k_x \mathbb{D}_y(F_2)(\mathbf{k})- k_y \mathbb{D}_x(F_2)(\mathbf{k})\right\},
\end{align*}
where now
\begin{align*}
    F_2(\mathbf{k})=\omega(k)A(\mathbf{k})e^{i\omega(k) t}.
\end{align*}
Thus we get 
\begin{align*}
 \mathcal{J}_2&\rightarrow -2i(2\pi)^2\int d\mathbf{k}\;\omega B(-\mathbf{k})(\mathbf{k}\cross\nabla_{\mathbf{k}})A(\mathbf{k}).
\end{align*}
Following the exact same approach we find that 
\begin{align*}
 \mathcal{J}_3&\rightarrow 2i(2\pi)^2\int d\mathbf{k}\;\omega A(-\mathbf{k})(\mathbf{k}\cross\nabla_{\mathbf{k}})B(\mathbf{k}),
\end{align*}
from which we conclude that the continuum limit for the total angular momentum $\mathcal{J}_L$ in the cavity is 
 \begin{align*}
 \mathcal{J}_L&\rightarrow -2i(2\pi)^2\int d\mathbf{k}\;\omega\left\{ B(-\mathbf{k})(\mathbf{k}\cross\nabla_{\mathbf{k}})A(\mathbf{k})-A(-\mathbf{k})(\mathbf{k}\cross\nabla_{\mathbf{k}})B(\mathbf{k})\right\}\nonumber\\
 &=2 \mathcal{J}.
\end{align*}
We almost get the correct continuum limit using our postulate; it is however too large by a factor of 2.

\subsubsection{A conjecture and a resolution of factor-of-two problem}
Defining $\tilde{\mathcal{J}}$ to be the exact continuum limit of the sum of $\mathcal{J}_2$ and $\mathcal{J}_3$, found using the exact same approach as on page \pageref{ExactContinuumLimitM1}, we have 
\begin{align*}
     &\tilde{\mathcal{J}}=-i(2\pi)^2\int d\mathbf{k}\big\{B(-\mathbf{k})e^{-i\omega(k)t}\left( k_x \mathbb{D}_y(F_2)(\mathbf{k})- k_y \mathbb{D}_x(F_2)(\mathbf{k})\right)\nonumber\\
     &-A(-\mathbf{k})e^{i\omega(k)t}\left( k_x \mathbb{D}_y(F_3)(\mathbf{k})- k_y \mathbb{D}_x(F_3)(\mathbf{k})\right)\big\},
\end{align*}
where
\begin{align*}
    F_2(\mathbf{k})&=\omega(k)A(\mathbf{k})e^{i\omega(k) t},\nonumber\\
    F_3(\mathbf{k})&=\omega(k)B(\mathbf{k})e^{-i\omega(k) t}.
\end{align*}
Recall that the total angular momentum for a scalar field in the continuum is
 \begin{align}
   \mathcal{J}&=-i(2\pi)^2\int d\mathbf{k}\;\omega\big\{B(-\mathbf{k})(\mathbf{k}\cross\nabla_{\mathbf{k}})A(\mathbf{k})\nonumber\\
   &-A(-\mathbf{k})(\mathbf{k}\cross\nabla_{\mathbf{k}})B(\mathbf{k})\big\}.\label{AngularMomentumContinuumSpectralFormula}
\end{align}
Since the operators $\mathbb{D}_x$ and $\mathbb{D}_y$ at this point are unresolved  formal limits of the extremely non-local operators $\mathbb{D}_x^L$ and $\mathbb{D}_y^L$, it is not at all evident that expression $\tilde{ \mathcal{J}}$ contains all the time independent terms, or that its  time dependent terms vanishes. At this point we know so little about the nature of the operators $\mathbb{D}_x$ and $\mathbb{D}_y$ that we just can't decide one way or the other.

There is however a strong formal similarity between the formulas for $\mathcal{J}$ and $\tilde{\mathcal{J}}$.
In fact, if we simply assume the correspondence
\begin{align*}
   \left( k_x \mathbb{D}_y(F_2)(\mathbf{k})- k_y \mathbb{D}_x(F_2)(\mathbf{k})\right)&=\omega (\mathbf{k}\cross\nabla_{\mathbf{k}})A(\mathbf{k})e^{i\omega(k) t},\nonumber\\
   \left( k_x \mathbb{D}_y(F_3)(\mathbf{k})- k_y \mathbb{D}_x(F_3)(\mathbf{k})\right)&=\omega (\mathbf{k}\cross\nabla_{\mathbf{k}})B(\mathbf{k})e^{-i\omega(k) t},
\end{align*}
then $\tilde{\mathcal{J}}=\mathcal{J}$, and the continuum limit of the total angular momentum of a scalar field in a square cavity will be exactly equal to the correct expression, with no factor of two appearing.  

From what we did in the previous section, it is also clear that if the correspondence holds, then the terms $\mathcal{J}_1$ and $\mathcal{J}_4$ vanish in the continuum limit.

Furthermore, it follows from our work in the previous subsection that the correspondence will be correct if the following conjecture holds
\begin{align}
    \mathbb{D}(f)(x)=f'(x),\label{conjecture}
\end{align}
where we recall from that section that 
\begin{align*}
    \mathbb{D}(f)(x)=\lim_{L\rightarrow\infty}\sum_{s=1}^{\infty}(-1)^{(s-1)}\frac{f(x+\triangle s)-f(x-\triangle s)}{\triangle s}.
\end{align*}

\section{Proving the conjecture}\label{ProvingConjecture}

Recall that the conjecture, as stated, is
\begin{align*}
    \mathbb{D}(f)(x)=f'(x).
\end{align*}

Note that a sufficient condition for the correctness of the spectral continuum formula (\ref{AngularMomentumContinuumSpectralFormula}), is that the spectral functions $A(\mathbf{k})$ and $B(\mathbf{k})$ are defined for all $\mathbf{k}\in \mathbb{R}^3$, and are continuously  differentiable.

The spectral functions are complex valued, but it is evident that we only need to prove the conjecture for real valued continuously differentiable function $f(x)$, defined on the whole real line.

Let us start by rewriting the $\mathbb{D}$ operator in the following way
\begin{align}
    \mathbb{D}(f)(x_0)&=\lim_{L\rightarrow\infty}\;\sum_{s=1}^{\infty}(-1)^{(s-1)}\frac{f(x_0+\triangle s)-f(x_0-\triangle s)}{\triangle s}\nonumber\\
    &=\lim_{\triangle\rightarrow 0}\;\sum_{s\neq 0}(-1)^{(s-1)}\frac{f(x_0+\triangle s)}{\triangle s}\nonumber\\
    &=\lim_{\triangle\rightarrow 0}\;\left\{\left(\sum_{s\neq 0}(-1)^{(s-1)}\frac{\delta(x-x_0-\triangle s)}{\triangle s}\right)(f)\right\}\nonumber\\
    &=\lim_{\triangle\rightarrow 0}\;\left\{ \mathbf{\mathcal{D}}(x-x_0)(f)\right\},\label{5.3.2}
\end{align}
where we have used the fact that $\triangle=\frac{2\pi}{L}$, and where we have defined a generalized function
\begin{align}
    \mathbf{\mathcal{D}}(x-x_0)=\sum_{s\neq 0}(-1)^{(s-1)}\frac{\delta(x-x_0-\triangle s)}{\triangle s}.\label{5.3.3}
\end{align}
Next, observe that if ${\comb}$ is the generalized function known as the {\it Dirac comb} , we have 
\begin{align}
    &\left(-\left(\frac{1}{\triangle}\sinc{\left(\frac{x-x_0}{\triangle}\right)}\right)'\comb{\left(\frac{x-x_0}{\triangle}\right)}\right)\nonumber\\
     &=-\frac{1}{\triangle^2}\sinc'{\left(\frac{x-x_0}{\triangle}\right)}\comb{\left(\frac{x-x_0}{\triangle}\right)}\nonumber\\
      &=-\frac{1}{\triangle^2}\left(\frac{\cos\pi\left(\frac{x-x_0}{\triangle}\right)}{\frac{x-x_0}{\triangle}}-\frac{\sin\pi\left(\frac{x-x_0}{\triangle}\right)}{\pi\left(\frac{x-x_0}{\triangle}\right)^2}\right)\comb{\left(\frac{x-x_0}{\triangle}\right)}\nonumber\\
      &=-\frac{1}{\triangle^2}\sum_{s\neq 0}\left(\frac{\cos s\pi}{s}-\frac{\sin s\pi }{\pi s^2}\right)\delta\left(\frac{x-x_0}{\triangle}-s\right)\nonumber\\
       &=\sum_{s\neq 0}(-1)^{s-1}\frac{\delta\left(\frac{x-x_0}{\triangle}-s\right)}{s\triangle^2}\nonumber\\
         &=\sum_{s\neq 0}(-1)^{s-1}\frac{\delta\left(x-x_0-\triangle s\right)}{s\triangle},\label{5.3.4}
\end{align}
where we have made use of the fact that
\begin{align*}
    \frac{\cos x\pi}{x}-\frac{\sin x\pi }{\pi x^2}\rightarrow 0,\;\text{when}\;x\rightarrow 0.
\end{align*}

Using  (\ref{5.3.3}) and (\ref{5.3.4}) we see that the generalized function  $\mathbf{\mathcal{D}}(x-x_0)$ can be written in the form

\begin{align}
    \mathbf{\mathcal{D}}(x-x_0)=-\left(\frac{1}{\triangle}\sinc{\left(\frac{x-x_0}{\triangle}\right)}\right)'\comb{\left(\frac{x-x_0}{\triangle}\right)}.\label{5.3.6}
\end{align}
We are now going to compute the Fourier transform of the generalized function (\ref{5.3.6}). For that purpose we note that in this paper we are using the following convention for the Fourier transform
\begin{align}
    F(k)&=\frac{1}{\sqrt{2\pi}}\int_{-\infty}^{\infty}dx\;f(x)\;e^{i k x},\nonumber\\
    f(x)&=\frac{1}{\sqrt{2\pi}}\int_{-\infty}^{\infty}dk\;F(k)\;e^{-i k x}.\label{5.3.7}
\end{align}
Using this convention, we have for any function, or generalized function, f, 
\begin{align*}
    f\left(\frac{x-x_0}{\triangle}\right)\;&\rightarrow\;\triangle e^{i k x_0}\;F(\triangle k),\nonumber\\
    f'\left(\frac{x-x_0}{\triangle}\right)\;&\rightarrow\;-i k \triangle^2 e^{i k x_0}\;F(\triangle k).
\end{align*}

Note that with our convention (\ref{5.3.7}), the Fourier transform of the Dirac $\comb$ is not equal to itself. We rather have 
\begin{align*}
    \comb(x)\rightarrow \frac{1}{\sqrt{2\pi}}\comb(k).
\end{align*}

From these simple facts we have
\begin{align}
    \mathbf{\mathcal{D}}(x-x_0)&\rightarrow -\left(\frac{1}{\triangle^2}\right)\left(-i k \triangle^2 e^{i k x_0}\;\rect(\triangle k)\right)*\left(\frac{\triangle}{\sqrt{2\pi}} e^{i k x_0}\;\comb(\triangle k)\right)\nonumber\\
    &=\left(i k e^{i k x_0}\rect(\triangle k)\right)*\left(\frac{\triangle}{\sqrt{2\pi}}  e^{i k x_0}\comb(\triangle k)\right)\nonumber\\
    &=\frac{i}{\sqrt{2\pi}}e^{i k x_0}\sum_{s=-\infty}^{\infty}(k-\frac{s}{\triangle})\rect(\triangle k -s),\label{5.3.11}
\end{align}
where we have used the fact that the Fourier transform of the $\sinc$ function is a rectangle
\begin{align*}
    \rect(k)=\left\{ \begin{aligned}
1,\;\;\; \abs{k}\leq\frac{1}{2}\\
0.\;\;\; \abs{k}>\frac{1}{2}\\
\end{aligned}  \right.
,
\end{align*}
and where $*$ is the convolution operator.

The Fourier transform of $\mathbf{\mathcal{D}}(x-x_0)$ takes for $x_0=0$ the form of a periodic piecewise linear function of slope equal one, and period $\frac{1}{\triangle}$. A classical {\it saw tooth} function.

\begin{figure}[h]
    \centering
    \includegraphics[width=0.8\linewidth]{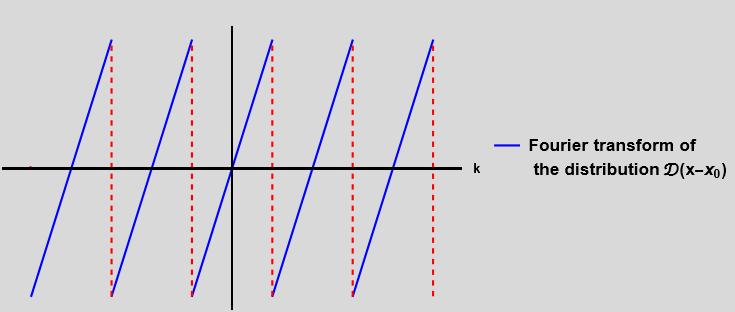}
    \caption{The real part of the Fourier transform of $\mathbf{\mathcal{D}}(x-x_0)$ for $x_0=0$ and $\triangle=1$. }
    \label{fig5.3.1}
\end{figure}
Using our convention for the Fourier transform, it is a simple matter to verify that the Fourier transform of the generalized function $-\delta'(x-x_0)$ is given by
\begin{align}
    -\delta'(x-x_0)\rightarrow \frac{ik}{\sqrt{2\pi}}e^{i k x_0}.\label{5.3.12}
\end{align}
Now, from (\ref{5.3.2}) we have for the Fourier transform of $\mathbb{D}(f)(x_0)$, 
\begin{align}
     \mathbb{D}(f)(x_0)&\rightarrow \lim_{\triangle\rightarrow 0}\;\left\{\left(\frac{i}{\sqrt{2\pi}}e^{i k x_0}\sum_{s=-\infty}^{\infty}(k-\frac{s}{\triangle})\rect(\triangle k -s)\right) (F)\right\}\nonumber\\
     &=\lim_{\triangle\rightarrow 0}\;\left\{\int_{-\infty}^{\infty}dk\; F(k)\frac{i}{\sqrt{2\pi}}e^{i k x_0}\sum_{s=-\infty}^{\infty}(k-\frac{s}{\triangle})\rect(\triangle k -s)\right\}\label{5.3.13}
\end{align}
\begin{figure}[h]
    \centering
    \includegraphics[width=0.8\linewidth]{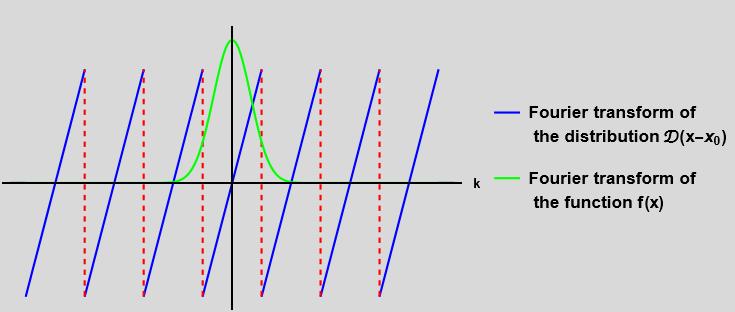}
    \caption{The real part of the integrand from formula (\ref{5.3.13}) for the case when $\triangle=1$, $x_0=0$ and the Fourier transform of f(x) is a Gaussian centered at $k=0$} 
    \label{fig5.3.2}
\end{figure}
Observe that when we take the limit when $\triangle\rightarrow 0$, the first period of the periodic piecewise linear function (\ref{5.3.11}), which is displayed in figure \ref{fig5.3.1}, will grow without bound. 

From this observation, together with identity (\ref{5.3.12}) and figure \ref{fig5.3.2}, we can now immediately conclude that
\begin{proposition}
 The conjecture is true for any function f(x) whose Fourier transform has compact support.  
\end{proposition}
Thus, compact support for the Fourier transform of a function $f(x)$ is a sufficient condition for the truth of the statement
\begin{align*}
    D(f)(x)=f'(x).
\end{align*}
This sufficient condition is however too strong for our purpse.  It is well known that functions whose Fourier transform has compact support are in fact analytic, and we need to prove the conjecture for the much larger class of continuously differentiable functions. The Fourier transform of such functions does not in general have compact support.

We will next show that the conjecture is also true for functions whose Fourier transform does not have compact support, but rather decay sufficiently fast for large $k$.
In order to show this we rewrite (\ref{5.3.13}) in the following way
\begin{align}
     \mathbb{D}(f)(x_0)&\rightarrow \lim_{\triangle\rightarrow 0}\;\left\{\int_{-\infty}^{\infty}dk\; F(k)\frac{i}{\sqrt{2\pi}}e^{i k x_0}\sum_{s=-\infty}^{\infty}(k-\frac{s}{\triangle})\rect(\triangle k -s)\right\}\nonumber\\
     &=\lim_{\triangle\rightarrow 0}\;\int_{-\infty}^{\infty}dk\; F(k)\frac{ik}{\sqrt{2\pi}}e^{i k x_0}\sum_{s=-\infty}^{\infty}\rect(\triangle k -s)\nonumber\\
     &-\lim_{\triangle\rightarrow 0}\;\int_{-\infty}^{\infty}dk\; F(k)\frac{i}{\sqrt{2\pi}}e^{i k x_0}\sum_{s=-\infty}^{\infty}\frac{s}{\triangle}\rect(\triangle k -s)\nonumber\\
      &=\int_{-\infty}^{\infty}dk\; F(k)\frac{ik}{\sqrt{2\pi}}e^{i k x_0}\nonumber\\
     &-\lim_{\triangle\rightarrow 0}\;\frac{i}{\sqrt{2\pi}}\sum_{s=-\infty}^{\infty}\int_{-\infty}^{\infty}dk\; F(k)e^{i k x_0}\frac{s}{\triangle}\rect(\triangle k -s).
     \label{5.3.15}
\end{align}
The first term in (\ref{5.3.15}) is, according to (\ref{5.3.12}), the Fourier transform of the generalized function $-\delta'(x-x_0)$, applied to the Fourier transform, $F(k)$, of our function $f(x)$. Thus, if the second term in (\ref{5.3.15}) is zero, we can conclude that
\begin{align}
    D(f)(x_0)=-\delta'(x-x_0)(f).\label{5.3.16}
\end{align}
Thus, the conjecture is true for functions for which the second term in (\ref{5.3.15}) is exactly zero. 

Observe that 
\begin{align}
    &\frac{i}{\sqrt{2\pi}}\sum_{s=-\infty}^{-1}\int_{-\infty}^{\infty}dk\; F(k)e^{i k x_0}\frac{s}{\triangle}\rect(\triangle k -s)\nonumber\\
    &=-\frac{i}{\sqrt{2\pi}}\sum_{s=1}^{\infty}\int_{-\infty}^{\infty}dk\; F(k)e^{i k x_0}\frac{s}{\triangle}\rect(\triangle k +s))\nonumber\\
    &=-\frac{i}{\sqrt{2\pi}}\sum_{s=1}^{\infty}\int_{-\infty}^{\infty}dk\; F(-k)e^{-ik x_0}\frac{s}{\triangle}\rect(-\triangle k +s))\nonumber\\  
    &=-\frac{i}{\sqrt{2\pi}}\sum_{s=1}^{\infty}\int_{-\infty}^{\infty}dk\; F(k)^*e^{-i k x_0}\frac{s}{\triangle}\rect(\triangle k -s))\nonumber\\ 
    &=\left\{\frac{i}{\sqrt{2\pi}}\sum_{s=1}^{\infty}\int_{-\infty}^{\infty}dk\; F(k)e^{i k x_0}\frac{s}{\triangle}\rect(\triangle k -s))\right\}^*.\label{5.3.17}
\end{align}
From (\ref{5.3.17}), (\ref{5.3.15}), the definition of convergence for a doubly infinite series, and the discussion above leading up to (\ref{5.3.16}), we can now conclude that
\begin{lemma}
    Let $f(x)$ be a function whose Fourier transform, $F(k)$, is a function, and let 
    \begin{align*}
        S(x_0,\triangle)=\sum_{s=1}^{\infty}\frac{s}{\triangle}\int_{\frac{s-\frac{1}{2}}{\triangle}}^{\frac{s+\frac{1}{2}}{\triangle}}dk\; F(k)e^{i k x_0}.
    \end{align*}
    
 Then 
    \begin{enumerate}
        \item The conjecture is true for $f(x)$ if 
        $\lim_{\triangle\rightarrow 0}\;S(x_0,\triangle)=0$.
        \item The conjecture is false for $f(x)$ if $\lim_{\triangle\rightarrow 0}\;S(x_0,\triangle)$ does not exist, or if the limit  exists but is nonzero.
    \end{enumerate}
   
\end{lemma}

\noindent This lemma tells us exactly what we need to do to verify that the conjecture is true for any particular $f(x)$.  
\vspace{3 mm}

%
\noindent  Let $f(x)$ be continuously differentiable. Then  it's Fourier transform, $F(k)$, satisfy the bound
    \begin{align*}
        \abs{F(k)}\leq\alpha\;k^{-\gamma},\;\;\;\text{for}\;k>c,\nonumber
    \end{align*}
    for some numbers $\alpha>0,c> 0$, and $\gamma>2$.
Given this, we have 
\begin{align}
     \abs{S(x_0,\triangle)}&\leq\alpha \sum_{s=1}^{\infty}\frac{s}{\triangle}\int_{\frac{s-\frac{1}{2}}{\triangle}}^{\frac{s+\frac{1}{2}}{\triangle}}dk\; k^{-\gamma}\nonumber\\
     &=\frac{\alpha}{\gamma-1} \sum_{s=1}^{\infty}\frac{s}{\triangle}\left(\left(\frac{s-\frac{1}{2}}{\triangle}\right)^{1-\gamma}-\left(\frac{s+\frac{1}{2}}{\triangle}\right)^{1-\gamma}\right)\nonumber\\
      &=\frac{\alpha\triangle^{\gamma-2}}{\gamma-1} \sum_{s=1}^{\infty}s\left(\left(s-\frac{1}{2}\right)^{1-\gamma}-\left(s+\frac{1}{2}\right)^{1-\gamma}\right),\label{5.3.19}
\end{align}
and since
\begin{align*}
    s\left(\left(s-\frac{1}{2}\right)^{1-\gamma}-\left(s+\frac{1}{2}\right)^{1-\gamma}\right)\sim (\gamma-1)\; s^{-(\gamma-1)},\;\;\;\text{when}\;\;s\rightarrow\infty,\nonumber
\end{align*}
the infinite series in (\ref{5.3.19}) converges. Given this, and the assumption that $\gamma>2$, it is now evident that
\begin{align*}
    \lim_{\triangle\rightarrow 0}\;S(x_0,\triangle)=0.\nonumber
\end{align*}
Thus we have 
\begin{theorem}
    Let $f(x)$ be continuously differentiable. Then
    \begin{align*}
         \mathbb{D}(f)(x)=f'(x),\;\;\;\text{for all}\; x\in\mathbb{R}
    \end{align*}
\end{theorem}

\section{Discussion}
For the physical problem described in the introduction, which was the motivation for this paper, Theorem 1 represents all we need to know about the $\mathbb{D}$-operator. 

However, beyond this context, exploring further properties of the $\mathbb{D}$-operator, and how it relates to differentiation, is certainly of mathematical interest.  We have not explored this topic deeply in the current paper, but based on extensive numerical calculations, some of which are displayed in Appendix E, and from comments from  scientists not directly involved in this work, we will suggest the following two conjectures on the extension of Theorem 1
\begin{conjecture}
     If $f(x)$ is continuous for all $x\in\mathbb{R}$, then 
     \begin{align*}
         \mathbb{D}(f)(x)=f'(x),
    \end{align*}
    for all $x\in\mathbb{R}$ where $f(x)$ is differentiable.
\end{conjecture}
 
\begin{conjecture}
    If $f(x)$ is discontinuous at one or more points in $\mathbb{R}$, then 
     \begin{align*}
         \mathbb{D}(f)(x)\neq f'(x),
    \end{align*}
    for all but a discrete set  of points.
\end{conjecture}
The discrete set  of points referred to in this second conjecture are typically points that have a symmetric position with respect to the points of discontinuity for the function $f(x)$. An example of this is the middle point of the hat function discussed on Appendix E.

\section*{Acknowledgment}
The authors are thankful for support from the Department of mathematics and statistics at the Arctic University of Norway, from the Arizona Center for Mathematical Sciences at the University of Arizona, and for the support from the Air Force Office for Scientific Research under Grant No. FA9550-19-1-0032. The authors are also thankful for insightful discussions and constructive critique of the ideas appearing in this paper from Miro Kolesik at the Wyant College of Optical Sciences, Arizona, USA.


\section*{Appendix A\label{appendixAA}}\nonumber
In this appendix we detail the derivation of the conserved quantities of linear and angular momentum for a 2D scalar field, which was left out of the main text.

\paragraph{Linear Momentum}
Consider an infinitesimal space translation along a direction $\vb{a} = (a,b)$, with $a^2 + b^2 = 1$, 
\begin{align*}
\vx = (x,y) \rightarrow \vx + \eps \; \vb{a}.
\end{align*}
This infinitesimal space translation induces a corresponding variation on scalar fields of the form 
\begin{align*}
u(t,x,y) \rightarrow u(t,x,y) + \eps(a \; u_x + b \; u_y). 
\end{align*}
Thus, for this case 
\begin{align*}
\eta(t,x,y)=a \; u_x + b \; u_y,
\end{align*}
and for this variation we have 
\begin{align*}
&\frac{\partial \mathcal{L}}{\partial u} \; \eta + \frac{\partial \mathcal{L}}{\partial u_t} \ \eta_t  + \frac{\partial \mathcal{L}}{\partial u_x} \; \eta_x+ \frac{\partial \mathcal{L}}{\partial u_y} \; \eta_y  \\ 
&=\frac{\partial \mathcal{L}}{\partial u_t} \; (a \; u_{xt} + b \; u_{yt} ) + \frac{\partial \mathcal{L}}{\partial u_x} \; (a \; u_{xx} + b \; u_{xy}) + \frac{\partial \mathcal{L}}{\partial u_y} \; (a \; u_{xy} + b \; u_{yy}) \nonumber \\
&= a \; (\frac{\partial \mathcal{L}}{\partial u_t} \; u_{xt} + \frac{\partial \mathcal{L}}{\partial u_x} \; u_{xx} + \frac{\partial \mathcal{L}}{\partial u_y} \; u_{xy}) \nonumber \\
&+ b \; (\frac{\partial  \mathcal{L}}{\partial u_t} \; u_{yt} + \frac{\partial \mathcal{L}}{\partial u_x} \; u_{xy} + \frac{\partial \mathcal{L}}{\partial u_y} \; u_{yy}) \nonumber \\ 
&= a \; \prt{x} \mathcal{L} + b \; \prt{y} \mathcal{L}, 
\end{align*}
so the functional is invariant with $F_1 = 0, \; F_2 = a \; \mathcal{L}, \; F_3 = b\; \mathcal{L}$.

The components of the Noether current are from (\ref{2.0.8}) 
\begin{align*}
j_1 &= - (a \; u_x + b \; u_y) \; u_t = - u_t \; (a \; u_x + b \; u_y),  \\
j_2 &=  \;a  (\inv{2}\; u_t^2 - \inv{2} \; c^2 \; u_x^2 - \inv{2} \; c^2 \; u_y^2) - (a \; u_x + b \; u_y) \; (-c^2 \; u_x) \nonumber \\ 
&= \inv{2} \;a \;  u_t^2 -  \inv{2} \;a \; c^2 \; u_x^2 - \inv{2} \;a \;  c^2 \; u_y^2 + a \; c^2 \; u_x^2 + b \; c^2 \; u_x \; u_y \nonumber \\
&= \inv{2} \; a \; u_t^2  + \inv{2} \; a \; c^2 \; u_x^2 - \inv{2} \; a \; c^2 \; u_y^2 + b \; c^2 \; u_x \; u_y, \nonumber \\
j_3 &= b\; (\inv{2} \; u_t^2 - \inv{2} \; c^2 \; u_x^2 - \inv{2} \; c^2 \; u_y^2) - (a \; u_x + b\; u_y) \; (-c^2 \; u_y) \nonumber \\ 
&= \inv{2} \; b \; u_t^2 - \inv{2} \; b \; c^2 \; u_x^2 + \inv{2} \; b \; c^2 \; u_y^2 + a \; c^2 \; u_x \; u_y, 
\end{align*}
and the conservation law is 
\begin{align*}
\prt{t} j_1 + \prt{x} j_2 + \prt{y} j_3 = 0.
\end{align*}
In order to simplify the conservation law we introduce a vector $\mathbf{p}$ and a Cartesian tensor $f_{\mathbf{p}}$ of rank 2, defined  by 
\begin{align*}
\mathbf{p}(t,\mathbf{x}) &= -u_t \; \grad{u}, \\
f_{\mathbf{p}}(t,\mathbf{x}) &= c^2 \; \grad{u} \; \grad{u} - \inv{2} \; c^2 \; \Tr (\grad{u} \; \grad{u}) \; I + \inv{2} \; u_t^2 \; I, \nonumber
\end{align*}
where $I$ is the identity matrix.

Using these quantities, we can write the local conservation law in the form
\begin{align}
\prt{t}( \vb{a} \vdot \mathbf{p}) + \div{(\vb{a} \vdot f_{\mathbf{p}})} =0.\label{2.2.8} 
\end{align}
The argument leading up to (\ref{2.2.8}) is true for all vectors $\vb{a}$. We therefore have the conservation law 
\begin{align}
\prt{t} \mathbf{p} + \div{f_{\mathbf{p}}} = 0. \label{2.2.9}
\end{align}
By definition, $\mathbf{p}$ is the linear momentum density and $f_{\mathbf{p}}$ is the linear momentum flux density. \\
The total linear momentum inside some domain $A$ is 
\begin{align*}
\mathcal{P} = \int_A d\mathbf{x}\; \mathbf{p}=  -\int_A d\mathbf{x}\; u_t \; \grad{u},
\end{align*}
and from (\ref{2.2.9}) we get 
\begin{align}
\frac{d \mathcal{P}}{dt} =& \int_A d\mathbf{x}\; \prt{t} \mathbf{p} = - \int_{S} dl \; f_{\mathbf{P}} \vdot \vb{n}\nonumber\\
=& - \int_{S} dl \;\left\{(\frac{1}{2}u_t^2-\frac{1}{2}c^2(\nabla u)^2)\mathbf{n}+c^2\nabla u\;\partial_{\mathbf{n}} u\right\}. \label{2.2.11} 
\end{align}

As was the case  for the energy, the linear momentum will clearly be conserved if  we assume that the domain D is so large, or the scalar field so localized, that it is negligible in a sufficiently small region around the boundary. If the field is not negligible near the boundary, neither  vanishing u, nor the flux of u, at the boundary, will ensure that the total momentum of the scalar field inside A is conserved.

\paragraph{Angular Momentum}
Consider an infinitesimal rotation of the form
\begin{align*}
    (x,y)\rightarrow (x,y)+\epsilon (-y,x).
\end{align*}
This infinitesimal rotation induces a variation on scalar fields of the form
\begin{align}
u(t,x,y)\rightarrow u(t,x,y)+\epsilon (xu_y-y u_x).\label{2.3.2}
\end{align}
Thus, for this case 
\begin{align*}
\eta(t,x,y)=xu_y-y u_x,
\end{align*}
and for this variation we have 
\begin{align*}
&\frac{\partial \mathcal{L}}{\partial u} \; \eta + \frac{\partial \mathcal{L}}{\partial u_t} \ \eta_t  + \frac{\partial \mathcal{L}}{\partial u_x} \; \eta_x+ \frac{\partial \mathcal{L}}{\partial u_y} \; \eta_y \nonumber \\ 
&=u_t(xu_{y}-yu_{x})_t-c^2u_x(xu_y-y u_x)_x-c^2(xu_y-y u_x)_y \nonumber \\
&=u_t(xu_{yt}-yu_{xt})-c^2u_x(u_y+xu_{xy}-yu_{xx})-c^2u_y(xu_{yy}-u_x-yu_{xy})\nonumber \\
&= (xu_tu_{yt}-yu_tu_{xt})-c^2(u_xu_y+xu_xu_{xy}-yu_xu_{xx})\nonumber\\
&-c^2(xu_yu_{yy}-u_yu_x-yu_yu_{xy})\nonumber \\ 
&=  xu_tu_{yt}-yu_tu_{xt}-c^2u_xu_y-c^2xu_xu_{xy}+c^2yu_xu_{xx}\nonumber\\
&-c^2xu_yu_{yy}+c^2u_yu_x+c^2yu_yu_{xy}\nonumber\\
&=\frac{1}{2}(xu_t^2)_y-\frac{1}{2}(yu_t^2)_x-c^2(\frac{1}{2}x(u_x^2+u_y^2))_y\nonumber\\
&+c^2(\frac{1}{2}y(u_x^2+u_y^2))_x\nonumber\\
&=\frac{1}{2}\left\{ -y u_t^2+c^2y(u_x^2+u_y^2) \right\}_x+\frac{1}{2}\left\{ x u_t^2-c^2x(u_x^2+u_y^2) \right\}_y. 
\end{align*}
So the functional is invariant with 
\begin{align}
    F_1&=0,\nonumber\\
    F_2&=\frac{1}{2}\left(-y u_t^2+c^2y(u_x^2+u_y^2)\right),\nonumber\\
    F_3&=\frac{1}{2}\left(x u_t^2-c^2x(u_x^2+u_y^2)\right).\label{2.3.5}
\end{align}
Using (\ref{2.3.2}) and (\ref{2.3.5}) in the formula for the components of the Noether current (\ref{2.0.8}), we find that
\begin{align*}
     j_1&=F_1-\eta\frac{\partial\mathcal{L}}{\partial_{u_t}}\nonumber\\
     &=-u_t(xu_y-yu_x),\nonumber\\
    j_2&=F_2-\eta\frac{\partial\mathcal{L}}{\partial_{u_x}}\nonumber\\
      &=-\frac{1}{2}yu_t^2-\frac{1}{2}c^2yu_x^2+\frac{1}{2}c^2yu_y^2+c^2xu_xu_y,\nonumber\\
    j_3&=F_3-\eta\frac{\partial\mathcal{L}}{\partial_{u_y}}\nonumber\\
     &=\frac{1}{2}xu_t^2-\frac{1}{2}c^2xu_x^2+\frac{1}{2}c^2xu_y^2-c^2yu_xu_y,
\end{align*}
and the local conservation law is 
\begin{align*}
\prt{t} j_1 + \prt{x} j_2 + \prt{y} j_3 = 0.
\end{align*}
As for energy and linear momentum, in order to simplify the local conservation law,  we introduce a scalar quantity $j$ and a vector quantity $f_j$ 
\begin{align*}
    j(t,\mathbf{x})=&-u_t\;\mathbf{x}\cross \nabla u,\nonumber\\
    f_j(t,\mathbf{x})=&\mathbf{x}^{\perp}\cdot f_{\mathbf{p}}.
\end{align*}
Note that the cross product of all plane vectors points in the $z$ direction; the scalar $m$ is here understood to be the $z$ component of the cross product $\mathbf{x}\cross \nabla u$. We have also here introduced the notation $\mathbf{x}^{\perp}=(-y,x)$.

Observe that the relation between linear momentum density  and angular momentum density
\begin{align*}
    j=\mathbf{x}\cross \mathbf{p},
\end{align*}
is the same as for the electromagnetic field, and is the one our intuition would lead us to expect.
Using these quantities, we can write the local conservation law for angular momentum in the form
\begin{align*}
\partial_t j+\nabla\cdot f_j=0.  
\end{align*}
By definition, $j$ is the angular momentum density and $f_j$ is the angular momentum flux density. \\
The total angular momentum inside some domain $A$ is 
\begin{align*}
\mathcal{J} = \int_A d\mathbf{x}\; j=  -\int_A d\mathbf{x}\; u_t\; \mathbf{x}\cross\nabla u.
\end{align*}
Like for the energy and the linear momentum, the angular momentum inside the domain $A$ is conserved if  we assume that the domain D is so large, or the scalar field so localized, that it is negligible in a sufficiently small region around the boundary. If the scalar field is not negligible in a small region around the boundary, the total angular momentum of the scalar field inside $A$ is in general not conserved. An exception to this general rule occur when $A$ takes the form of a circular disk. For this case the angular momentum is conserved if, on any section of the boundary, either $u=$const or $\partial_{\mathbf{n}}u=0$. 

\section*{Appendix B\label{appendixB}}
In this appendix we detail calculations relating to the continuum limit of energy, which are left out of the main text.

\paragraph{The continuum}

 Rewrite (\ref{2.1.10.1})  into the form
\begin{align}
    \mathcal{E}=\mathcal{E}_1+\mathcal{E}_2,\label{3.1.1.1}
\end{align}
where
\begin{align}
    \mathcal{E}_1&=\frac{1}{2}\int d\mathbf{x}\;u_t^2,\nonumber\\
    \mathcal{E}_2&=\frac{1}{2}c^2\int d\mathbf{x}\;(\nabla u)^2.\label{3.1.1.2}
\end{align}

From these first of the expressions (\ref{3.1.1.3}), we get 
\begin{align}
    u_t^2&=\int d\mathbf{k}\;d\mathbf{k}'\;(-\omega\omega')e^{i(\mathbf{k}+\mathbf{k}')\cdot\mathbf{x}}\nonumber\\
    &\left\{Ae^{-i\omega t}-Be^{i\omega t}\right\}\left\{A'e^{-i\omega' t}-B'e^{i\omega' t}\right\}\nonumber\\
    &=\int d\mathbf{k}\;d\mathbf{k}'\;(-\omega\omega')e^{i(\mathbf{k}+\mathbf{k}')\cdot\mathbf{x}}\nonumber\\
    &\left\{ AA'e^{-i(\omega+\omega') t} - AB'e^{-i(\omega-\omega') t}-BA'e^{i(\omega-\omega') t} + BB'e^{i(\omega+\omega') t}  \right\}.\label{3.1.1.4}
\end{align}
Here we use the compact notation $A'=A(\mathbf{k}')$, $\omega'=\omega(k')$, etc.

From the second of the two expressions (\ref{3.1.1.3}) we get in a similar way 
\begin{align}
    (\nabla u)^2&=\int d\mathbf{k}\;d\mathbf{k}'\;(-\mathbf{k}\mathbf{k}')e^{i(\mathbf{k}+\mathbf{k}')\cdot\mathbf{x}}\nonumber\\
     &\left\{Ae^{-i\omega t}+Be^{i\omega t}\right\}\left\{A'e^{-i\omega' t}+B'e^{i\omega' t}\right\}\nonumber\\
      &=\int d\mathbf{k}\;d\mathbf{k}'\;(-\mathbf{k}\mathbf{k}')e^{i(\mathbf{k}+\mathbf{k}')\cdot\mathbf{x}}\nonumber\\
    &\left\{ AA'e^{-i(\omega+\omega') t} + AB'e^{-i(\omega-\omega') t}+BA'e^{i(\omega-\omega') t} + BB'e^{i(\omega+\omega') t}  \right\}.\label{3.1.1.5}
\end{align}
Inserting  the expressions (\ref{3.1.1.4}) and (\ref{3.1.1.5}) into the formulas (\ref{3.1.1.2}), and performing the integration over physical space, we get
\begin{align}
    \mathcal{E}_1&=\frac{1}{2}(2\pi)^2\int d\mathbf{k}\;d\mathbf{k}'\;(-\omega\omega')\delta(\mathbf{k}+\mathbf{k}')\nonumber\\
    &\left\{ AA'e^{-i(\omega+\omega') t} - AB'e^{-i(\omega-\omega') t}-BA'e^{i(\omega-\omega') t} + BB'e^{i(\omega+\omega') t}  \right\}\nonumber\\
    &=\frac{1}{2}(2\pi)^2\int d\mathbf{k}\;\omega^2\left\{\abs{A(\mathbf{k})}^2+ \abs{B(\mathbf{k})}^2 \right\}\nonumber\\
    &-\frac{1}{2}(2\pi)^2\int d\mathbf{k}\;\omega^2A(\mathbf{k})B(\mathbf{k})^*e^{-2 i\omega t}\nonumber\\
    &-\frac{1}{2}(2\pi)^2\int d\mathbf{k}\;\omega^2B(\mathbf{k})A(\mathbf{k})^*e^{2 i\omega t},\label{3.1.1.6}
\end{align}
and
\begin{align*}
    \mathcal{E}_2&=\frac{1}{2}c^2(2\pi)^2\int d\mathbf{k}\;d\mathbf{k}'\;(-\mathbf{k}\mathbf{k}')\delta(\mathbf{k}+\mathbf{k}')\nonumber\\
    &\left\{ AA'e^{-i(\omega+\omega') t} +AB'e^{-i(\omega-\omega') t}+BA'e^{i(\omega-\omega') t} + BB'e^{i(\omega+\omega') t}  \right\}\nonumber\\
    &=\frac{1}{2}(2\pi)^2\int d\mathbf{k}\;c^2k^2\left\{\abs{A(\mathbf{k})}^2+ \abs{B(\mathbf{k})}^2 \right\}\nonumber\\
    &+\frac{1}{2}(2\pi)^2\int d\mathbf{k}\;c^2k^2A(\mathbf{k})B(\mathbf{k})^*e^{-2 i\omega t}\nonumber\\
    &+\frac{1}{2}(2\pi)^2\int d\mathbf{k}\;c^2k^2B(\mathbf{k})A(\mathbf{k})^*e^{2 i\omega t}.
\end{align*}

Finally inserting (\ref{3.1.1.5}) and (\ref{3.1.1.6}) into the expression for the energy (\ref{3.1.1.1}), we get
\begin{align*}
    \mathcal{E}&=\mathcal{E}_1+\mathcal{E}_2\nonumber\\
    &=\frac{1}{2}(2\pi)^2\int d\mathbf{k}\;(\omega^2+c^2k^2)\left\{\abs{A(\mathbf{k})}^2+ \abs{B(\mathbf{k})}^2 \right\}\nonumber\\
    &+\frac{1}{2}(2\pi)^2\int d\mathbf{k}\;(-\omega^2+c^2k^2)A(\mathbf{k})B(\mathbf{k})^*e^{-2 i\omega t}\nonumber\\
     &+\frac{1}{2}(2\pi)^2\int d\mathbf{k}\;(-\omega^2+c^2k^2)B(\mathbf{k})A(\mathbf{k})^*e^{2 i\omega t}\nonumber\\
      &=(2\pi)^2\int d\mathbf{k}\;\omega^2\left\{\abs{A(\mathbf{k})}^2+ \abs{B(\mathbf{k})}^2 \right\},
\end{align*}
where in the last step we used the dispersion relation. 

\paragraph{A square cavity}

We proceed like for the continuum case by splitting the space-time formula for the energy into the sum of two pieces
\begin{align*}
    \mathcal{E}_L= \mathcal{E}_1+ \mathcal{E}_2,
\end{align*}
where
\begin{align}
     \mathcal{E}_1&=\frac{1}{2}\int_A d\mathbf{x}\;u_t^2,\nonumber\\
     \mathcal{E}_2&=\frac{1}{2}c^2\int_A d\mathbf{x}\;(\nabla u)^2.\label{3.2.1.2}
\end{align}
These integrals now go over the cavity only, not the whole physical space, like for the continuum case.

Inserting  the formulas (\ref{3.2.1.3})  into the formula for $ \mathcal{E}_1$ from (\ref{3.2.1.2}) we have
\begin{align}
     \mathcal{E}_1&=-\frac{1}{2}\int_Ad\mathbf{x}\sum_{\mathbf{n}\mathbf{n}'}\;\omega\omega'e^{i (\mathbf{k}+\mathbf{k}')\cdot\mathbf{x}}\nonumber\\
    &\left\{ ae^{-i\omega t}-  be^{i\omega t}\right\}\left\{ a'e^{-i\omega' t}-  b'e^{i\omega't}\right\}\nonumber\\
    &=-\frac{1}{2}\sum_{\mathbf{n}\mathbf{n}'}\;\omega\omega'S_{\mathbf{n}\mathbf{n}'}\nonumber\\
    &\left\{aa'e^{-i(\omega+\omega')t}-ab'e^{-i(\omega-\omega')t}-ba'e^{i(\omega-\omega')t}+bb'e^{i(\omega+\omega')t} \right\},\label{3.2.1.4}
    \end{align}
    where we are using the compact notation $a=a_{\mathbf{n}},a'=a_{\mathbf{n}'}$, etc.

    The quantity $S_{\mathbf{n}\mathbf{n}'}$ is by definition
    \begin{align}
        S_{\mathbf{n}\mathbf{n}'}&=\int_A d\mathbf{x}\;e^{i (\mathbf{k}_{\mathbf{n}}+\mathbf{k}_{\mathbf{n}'})}\nonumber\\
        &=L^2\delta_{\mathbf{n}-\mathbf{n}'},\label{3.2.1.5}
    \end{align}
    where $\delta_{\mathbf{p}}$ is the Kronecker delta.

    Inserting (\ref{3.2.1.5}) into the expression for the kinetic energy (\ref{3.2.1.4}) we get
    \begin{align}
         \mathcal{E}_1&=\frac{L^2}{2}\sum_{\mathbf{n}}\;\omega_{n}^2\left\{\abs{a_{\mathbf{n}}}^2+\abs{b_{\mathbf{n}}}^2\right\}\nonumber\\
        &-\frac{L^2}{2}\sum_{\mathbf{n}}\;\omega_{n}^2a_{\mathbf{n}}a_{-\mathbf{n}}e^{-2 i\omega_{n}}\nonumber\\
         &-\frac{L^2}{2}\sum_{\mathbf{n}}\;\omega_{n}^2b_{\mathbf{n}}b_{-\mathbf{n}}e^{2 i\omega_{n}}.\label{3.2.1.6}
    \end{align}
    An entirely similar calculation leads to the following formula for $ \mathcal{E}_2$
    \begin{align}
       \mathcal{E}_2&=\frac{L^2}{2}\sum_{\mathbf{n}}\;c^2\mathbf{k}_{\mathbf{n}}^2\left\{\abs{a_{\mathbf{n}}}^2+\abs{b_{\mathbf{n}}}^2\right\}\nonumber\\
        &+\frac{L^2}{2}\sum_{\mathbf{n}}\;c^2\mathbf{k}_{\mathbf{n}}^2a_{\mathbf{n}}a_{-\mathbf{n}}e^{-2 i\omega_{n}t}\nonumber\\
         &+\frac{L^2}{2}\sum_{\mathbf{n}}\;c^2\mathbf{k}_{\mathbf{n}}^2b_{\mathbf{n}}b_{-\mathbf{n}}e^{2 i\omega_{n}t}.\label{3.2.1.7}   
    \end{align}
    Adding together the expressions for $ \mathcal{E}_1$ and $ \mathcal{E}_2$, (\ref{3.2.1.6}) and (\ref{3.2.1.7}), and using the dispersion relation (\ref{3.2.4}), we get the following simple expression of the total energy, of the scalar field in a square cavity, expressed in terms of discrete cavity modes
    \begin{align*}
         \mathcal{E}_L=L^2\sum_{\mathbf{n}}\;\omega_{n}^2\left\{\abs{a_{\mathbf{n}}}^2+\abs{b_{\mathbf{n}}}^2\right\}.
    \end{align*}

\section*{Appendix C\label{appendixC}}
In this appendix we detail calculations relating to the continuum limit of linear momentum, which are left out of the main text.

\paragraph{The continuum}
Using expressions (\ref{3.1.1.3}) and the expression for the total linear momentum of the scalar field (\ref{2.2.10}), we have 
\begin{align}
    \mathcal{P}&=-\int d\mathbf{x}\;u_t\nabla U\nonumber\\
    &=-\int d\mathbf{x}\int d\mathbf{k} d\mathbf{k}'\omega\mathbf{k}'\;e^{i(\mathbf{k}+\mathbf{k}')\cdot\mathbf{x}}\nonumber\\
    &\left\{A e^{-i\omega t}-Be^{i\omega t}\right\}\left\{A' e^{-i\omega' t}+B' e^{i\omega' t}\right\}\nonumber\\
    &=-(2\pi)^2\int d\mathbf{k} d\mathbf{k}'\omega\mathbf{k}'\;\delta(\mathbf{k}+\mathbf{k}')\nonumber\\
    &\left\{AA'e^{-i(\omega+\omega')t}+AB'e^{-i(\omega-\omega') t}-BA'e^{i(\omega-\omega') t}-BB'e^{i(\omega+\omega') t}\right\}\nonumber\\
    &=(2\pi)^2\int d\mathbf{k}\;\omega\left\{\mathbf{k}\abs{A(\mathbf{k})}^2-\mathbf{k}\abs{B(\mathbf{k})}^2 \right\}\nonumber\\
     &\;\;\;\;+(2\pi)^2\int d\mathbf{k}\;\omega\mathbf{k}A(\mathbf{k})A(-\mathbf{k})e^{-2 i\omega t}\nonumber\\
      &\;\;\;\;-(2\pi)^2\int d\mathbf{k}\;\omega\mathbf{k}B(\mathbf{k})B(-\mathbf{k})e^{2 i\omega t}.
     \label{3.1.2.1}
\end{align}
Observe that the integrand in each of last two time dependent terms of (\ref{3.1.2.1}) is an odd function of $\mathbf{k}$, when restricted to  circles $\omega(k)=$const. From this it follows that both those terms vanish.

Thus, the formula for total linear momentum  of the scalar field, expressed in terms of spectral amplitudes,  is time invariant, as it should be, and takes the form
\begin{align*}
    \mathcal{P}&= (2\pi)^2\int d\mathbf{k}\;\omega\left\{\mathbf{k}\abs{A(\mathbf{k})}^2-\mathbf{k}\abs{B(\mathbf{k})}^2 \right\}.
\end{align*}
As for the total energy, we  note that the total linear momentum in the scalar field is simply a sum over the individual momenta in all the plane waves that are used to construct the field. This is again what  intuition leads us to expect.

\paragraph{A square cavity}
 Using the expressions (\ref{3.2.1.3}) we have
\begin{align}
    \mathcal{P}_L&=-\int_Ad\mathbf{x}\;u_t\nabla u\nonumber\\
    &=-\int_A d\mathbf{x}\sum_{\mathbf{n}\mathbf{n}'}\;\omega\mathbf{k}'e^{i (\mathbf{k}+\mathbf{k}')\cdot\mathbf{x}}\nonumber\\
    &\left\{ ae^{-i\omega t}-  be^{i\omega t}\right\}\left\{ a'e^{-i\omega' t}+  b'e^{i\omega't}\right\}
\nonumber\\
&=-\sum_{\mathbf{n}\mathbf{n}'}\;\omega\mathbf{k}'L^2\delta_{\mathbf{n}-\mathbf{n}'}\nonumber\\
&\left\{aa'e^{-i(\omega+\omega')t}+ab'e^{-i(\omega-\omega')t}-ba'e^{i(\omega-\omega')t}-bb'e^{i(\omega+\omega')t} \right\}\nonumber\\
&=L^2\sum_{\mathbf{n}}\;\omega_{n}\left\{\mathbf{k}_{\mathbf{n}}\abs{a_{\mathbf{n}}}^2-\mathbf{k}_{\mathbf{n}}\abs{b_{\mathbf{n}}}^2\right\}\nonumber\\
        &+L^2\sum_{\mathbf{n}}\;\omega_{n}\mathbf{k}_{\mathbf{n}}a_{\mathbf{n}}a_{-\mathbf{n}}e^{-2 i\omega_{n}t}\nonumber\\
         &-L^2\sum_{\mathbf{n}}\;\omega_{n}\mathbf{k}_{\mathbf{n}}b_{\mathbf{n}}b_{-\mathbf{n}}e^{2 i\omega_{n}t}.\label{3.2.2.1}
\end{align}
Note that when $\mathbf{n}\rightarrow -\mathbf{n}$
\begin{align}
   \omega_{n}\mathbf{k}_{\mathbf{n}}a_{\mathbf{n}}a_{-\mathbf{n}}e^{-2 i\omega_{n}t}\rightarrow\omega_{n}\mathbf{k}_{-\mathbf{n}}a_{-\mathbf{n}}a_{\mathbf{n}}e^{-2 i\omega_{n}t}=-\omega_{n}\mathbf{k}_{\mathbf{n}}a_{\mathbf{n}}a_{-\mathbf{n}}e^{-2 i\omega_{n}t}.\label{3.2.2.2}
\end{align}
Because of this, the second, time dependent term in (\ref{3.2.2.1}), vanishes. For a similar reason, the third, time dependent term in (\ref{3.2.2.1}), also vanishes.

Thus, we get the following simple expression of the total linear momentum, of the scalar field in a square cavity, expressed in terms of discrete cavity modes
\begin{align*}
    \mathcal{P}_L= L^2\sum_{\mathbf{n}}\;\omega_{n}\left\{\mathbf{k}_{\mathbf{n}}\abs{a_{\mathbf{n}}}^2-\mathbf{k}_{\mathbf{n}}\abs{b_{\mathbf{n}}}^2\right\}.
\end{align*}

\section*{Appendix D\label{appendixD}}
In this appendix we detail calculations relating to the continuum limit of angular momentum, which are  left out of the main text.

\paragraph{The continuum}
Using expressions (\ref{3.1.1.3}) and the expression for the total angular momentum of the scalar field (\ref{2.3.11}), we have 
\begin{align}
    \mathcal{J}&=  -\int d\mathbf{x}\; u_t\; \mathbf{x}\cross\nabla u \\
    &=-\int d\mathbf{x}\int d\mathbf{k} d\mathbf{k}'\omega'\;(\mathbf{x}\cross\mathbf{k})\;e^{i(\mathbf{k}+\mathbf{k}')\cdot\mathbf{x}}\nonumber\\
    &\left\{A' e^{-i\omega' t}-B'e^{i\omega' t}\right\}\left\{A e^{-i\omega t}+B e^{i\omega t}\right\}\nonumber\\
    &=-\int d\mathbf{k} d\mathbf{k}'\omega'\;S(\mathbf{k},\mathbf{k}')\nonumber\\
    &\left\{A'Ae^{-i(\omega'+\omega)t}+A'Be^{-i(\omega'-\omega) t}-B'Ae^{i(\omega'-\omega) t}-B'Be^{i(\omega'+\omega) t}\right\},
     \label{3.1.3.1}
\end{align}
where 
\begin{align}
  S(\mathbf{k},\mathbf{k}')&= \int d\mathbf{x}\;(\mathbf{x}\cross\mathbf{k})e^{i(\mathbf{k}+\mathbf{k}')\cdot\mathbf{x}}\nonumber\\
  &=\int d\mathbf{x}\;(xk_y-yk_x)e^{i(\mathbf{k}+\mathbf{k}')\cdot\mathbf{x}}\nonumber\\
   &=-i\int d\mathbf{x}\;(k_y\partial_{k_x}-k_x\partial_{k_y})e^{i(\mathbf{k}+\mathbf{k}')\cdot\mathbf{x}}\nonumber\\
   &=i\;(\mathbf{k}\cross\nabla_{\mathbf{k}}) \int d\mathbf{x}\; e^{i(\mathbf{k}+\mathbf{k}')\cdot\mathbf{x}}\nonumber\\ 
   &=i(2\pi)^2\;(\mathbf{k}\cross\nabla_{\mathbf{k}})\delta(\mathbf{k}+\mathbf{k}').\label{3.1.3.2}
\end{align}
Inserting (\ref{3.1.3.2}) into (\ref{3.1.3.1}) we get
\begin{align*}
  \mathcal{J}&=-i(2\pi)^2\int d\mathbf{k} d\mathbf{k}'\omega'(\mathbf{k}\cross\nabla_{\mathbf{k}})\delta(\mathbf{k}+\mathbf{k}')\;\nonumber\\
    &\left\{A'Ae^{-i(\omega'+\omega)t}+A'Be^{-i(\omega'-\omega) t}-B'Ae^{i(\omega'-\omega) t}-B'Be^{i(\omega'+\omega) t}\right\}\nonumber\\
    &=i(2\pi)^2\int d\mathbf{k} d\mathbf{k}'\omega'\delta(\mathbf{k}+\mathbf{k}')(\mathbf{k}\cross\nabla_{\mathbf{k}})\;\nonumber\\
    &\left\{A'Ae^{-i(\omega'+\omega)t}+A'Be^{-i(\omega'-\omega) t}-B'Ae^{i(\omega'-\omega) t}-B'Be^{i(\omega'+\omega) t}\right\},
\end{align*}
where we have used the rules for distributional derivatives, and the fact that the differential operator commutes with functions of $\mathbf{k}'$.
Thus 
\begin{align*}
    \omega(\mathbf{k}')(\mathbf{k}\cross\nabla_{\mathbf{k}})=(\mathbf{k}\cross\nabla_{\mathbf{k}})\omega(\mathbf{k}').
\end{align*}
The formula for $ \mathcal{J}$ now simplifies into
\begin{align}
    \mathcal{J}&=-i(2\pi)^2\int d\mathbf{k}\;\omega\left\{B(-\mathbf{k})(\mathbf{k}\cross\nabla_{\mathbf{k}})A(\mathbf{k})-A(-\mathbf{k})(\mathbf{k}\cross\nabla_{\mathbf{k}})B(\mathbf{k})\right\}\nonumber\\
    &+i(2\pi)^2\int d\mathbf{k}\;\omega\;A(-\mathbf{k})(\mathbf{k}\cross\nabla_{\mathbf{k}})A(\mathbf{k})e^{-2i\omega t}\nonumber\\
   &-i(2\pi)^2\int d\mathbf{k}\;\omega\;B(-\mathbf{k})(\mathbf{k}\cross\nabla_{\mathbf{k}})B(\mathbf{k})e^{2i\omega t}.\label{3.1.3.4}
\end{align}
The integrand of the last two terms are time dependent, and since the total angular momentum of the scalar field is conserved, they must both cancel. That they do so is not immediately obvious, but observe that for the first of the two time dependent terms in (\ref{3.1.3.4}), we have
\begin{align*}
    &i(2\pi)^2\int d\mathbf{k}\;\omega\;A(-\mathbf{k})(\mathbf{k}\cross\nabla_{\mathbf{k}})A(\mathbf{k})e^{-2i\omega t}\nonumber\\
    &=\frac{1}{2}i(2\pi)^2\int d\mathbf{k}\;\omega\;\left\{A(-\mathbf{k})(\mathbf{k}\cross\nabla_{\mathbf{k}})A(\mathbf{k})+A(-\mathbf{k})(\mathbf{k}\cross\nabla_{\mathbf{k}})A(\mathbf{k})\right\}e^{-2i\omega t}\nonumber\\
    &=\frac{1}{2}i(2\pi)^2\int d\mathbf{k}\;\omega\;\left\{A(-\mathbf{k})(\mathbf{k}\cross\nabla_{\mathbf{k}})A(\mathbf{k})+A(\mathbf{k})(\mathbf{k}\cross\nabla_{\mathbf{k}})A(-\mathbf{k})\right\}e^{-2i\omega t}\nonumber\\
     &=\frac{1}{2}i(2\pi)^2\int d\mathbf{k}\;\omega\;(\mathbf{k}\cross\nabla_{\mathbf{k}})(A(\mathbf{k})A(-\mathbf{k}))e^{-2i\omega t}.
\end{align*}
Note that on each circle $C$, defined by $\omega(k)=$const, we end up having to solve  an integral of the form  
\begin{align*}
    \int_Cdl\;\mathbf{k}\cross\nabla_{\mathbf{k}}\phi,
\end{align*}
where $\phi$ is a scalar function.
\label{VanishingIntegral}
We conclude, from a  version of Stokes theorem\cite{Jones,Per}, that this integral is zero. One can also show this fact explicitly by doing the integral using polar coordinates. Thus, the last two terms in (\ref{3.1.3.4}) both cancel.

The formula for total angular momentum  of the scalar field, expressed in terms of spectral amplitudes, becomes time invariant, as it should be, and takes the form
\begin{align*}
    \mathcal{J}&=-i(2\pi)^2\int d\mathbf{k}\;\omega\left\{A(\mathbf{k})^*(\mathbf{k}\cross\nabla_{\mathbf{k}})A(\mathbf{k})-B(\mathbf{k})^*(\mathbf{k}\cross\nabla_{\mathbf{k}})B(\mathbf{k})\right\}.
\end{align*}

\paragraph{A square cavity}
Observe that
\begin{align*}
  S_{\mathbf{n}'\mathbf{n}}= iL^2\left(n_x\delta_{n_x+n_x'}\frac{(-1)^{\abs{n_y+n_y'}}}{n_y+n_y'}-n_y\delta_{n_y+n_y'}\frac{(-1)^{\abs{n_x+n_x'}}}{n_x+n_x'}\right),
\end{align*}
 can be nonzero only when
\begin{align*}
    n_x'=-n_x,\;\;\;\;\text{and}\;\;\;n_y'\neq-n_y,
\end{align*}
or if
\begin{align*}
    n_y'=-n_y,\;\;\;\;\text{and}\;\;\;n_x'\neq-n_x.
\end{align*}
For these two cases, $S_{\mathbf{n}'\mathbf{n}}$ takes the form
\begin{align*}
    S_{\mathbf{n}'\mathbf{n}}&=iL^2n_xP_{n_y+n_y'},\;\;\;\mathbf{n}=(n_x,n_y),\mathbf{n}'=(-n_x,n_y'),\nonumber\\
     S_{\mathbf{n}'\mathbf{n}}&=-iL^2n_yP_{n_x+n_x'},\;\;\;\mathbf{n}=(n_x,n_y),\mathbf{n}'=(n_x',-n_y),
\end{align*}
where
\begin{align*}
    P_q&=\frac{(-1)^{\abs{q}}}{q},\;\;\;q\neq 0,\nonumber\\
    P_0&=0.
\end{align*}
For the first contribution to the total angular momentum in the cavity, $\mathcal{J}_1$, we have 
\begin{align*}
   \mathcal{J}_1&=-iL^2\sum_{n_x,n_y,n_y'}\;n_xP_{n_y+n_y'}\omega_{-n_x,n_y'} a_{-n_x,n_y'}a_{n_x,n_y}e^{-i(\omega_{n_x,n_y}+\omega_{-nx,ny'})t}\nonumber\\
  &+iL^2\sum_{n_x,n_x',n_y}\;n_yP_{n_x+n_x'}\omega_{n_x',-n_y} a_{n_x',-n_y}a_{n_x,n_y}e^{-i(\omega_{n_x,n_y}+\omega_{n_x',-n_y})t}.
\end{align*}
Note that in this formula, and in the ones to follow,  we have introduced the notation
\begin{align*}
    \omega_{p,q}=\frac{2\pi c}{L}(p^2+q^2)^{\frac{1}{2}}.
\end{align*}
We evidently have that
\begin{align*}
    \omega_{-p,q}=\omega_{p,-q}=\omega_{p,q}.
\end{align*}
We will rewrite formula $ \mathcal{J}_1$ into a form that will be more convenient when  we consider the continuum limit, $L\rightarrow\infty$, in section \ref{AngularMomentumContinuumLimit}.
\begin{align*}
    & \mathcal{J}_1=-iL^2\sum_{n_x,n_y}a_{n_x,n_y}e^{-i\omega_{n_x,n_y}}\{n_x\sum_{n_y'}P_{n_y+n_y'}\omega_{n_x,n_y'}a_{-n_x,n_y'}e^{-i\omega_{n_x,n_y'}t}\nonumber\\
    &-n_y\sum_{n_x'}P_{n_x+n_x'}\omega_{n_x',n_y}a_{n_x',-n_y}e^{-i\omega_{n_x',n_y}t}\}\nonumber\\
    \nonumber\\
    &n_y+n_y'=u,\;n_x+n_x'=v\nonumber\\
    \nonumber\\
    &=-i L^2\sum_{n_x,n_y}a_{n_x,n_y}e^{-i\omega_{n_x,n_y}}\{n_x\sum_uP_u\omega_{n_x,-n_y+u}a_{-n_x,-n_y+u}e^{-i\omega_{n_x,-n_y+u}t}\nonumber\\
    &-n_y\sum_vP_v\omega_{-n_x+v,n_y}a_{-n_x+v,-n_y}e^{-i\omega_{-n_x+v,n_y}t}\}\nonumber\\
     \\
    &n_x\rightarrow-n_x,\;n_y\rightarrow-n_y\nonumber\\
    \nonumber\\
     &=i L^2\sum_{n_x,n_y}a_{-n_x,-n_y}e^{-i\omega_{n_x,n_y}}\{n_x\sum_uP_u\omega_{n_x,n_y+u}a_{n_x,n_y+u}e^{-i\omega_{n_x,n_y+u}t}\nonumber\\
     &-n_y\sum_v P_v\omega_{n_x+v,n_y}a_{n_x+v,n_y}e^{-i\omega_{n_x+v,n_y}t}\}.
\end{align*}
For any sequence $\alpha_\mathbf{n}=\alpha_{n_x,n_y}$  define two operators
\begin{align*}
    \Lambda_x(\alpha)_{nx,ny}&=\sum_s\;P_s \alpha_{n_x+s,n_y},\nonumber\\
    \Lambda_y(\alpha)_{n_x,n_y}&=\sum_s\;P_s \alpha_{n_x,n_y+s}.
\end{align*}
Given these two operators, we can write $\mathcal{J}_1$ in the form
\begin{align*}
     \mathcal{J}_1&=i L^2\sum_{n_x,n_y}a_{-n_x,-n_y}e^{-i\omega_{n_x,n_y}}\left\{ n_x \Lambda_y(f_1)_{n_x,n_y}-n_y\Lambda_x(f_1)_{n_x,n_y}\right\}\nonumber\\
    &=i L^2\sum_{\mathbf{n}}a_{-\mathbf{n}}e^{-i\omega_{n}t}\left\{ n_x \Lambda_y(f_1)_{\mathbf{n}}-n_y\Lambda_x(f_1)_{\mathbf{n}}\right\},
\end{align*}
where
\begin{align*}
    (f_1)_{\mathbf{n}}=\omega_{n}a_{\mathbf{n}}e^{-i\omega_{n}t}.
\end{align*}

\section*{Appendix E\label{appendixE}}

In this appendix we present numerical calculations of the  $\mathbb{D}(f)$ for some select functions, $f$. Some of these functions are covered by Theorem 1 from section \ref{ProvingConjecture}, and can thus our calculations can be though of as a sanity check on the proof of that theorem. Other functions are not covered by the Theorem, and our numerical calculations can be considered to be a first stab at figuring out if Theorem 1 holds true for a class of functions larger than the continuously differentiable ones.

In this numerical section, for any function $f(x)$ we consider, we evidently cannot take the limit as $L$ approaches infinity. This is what is required in order to calculate $\mathbb{D}(f)$. In order to not confuse the presentation, we introduce here the notation
\begin{align*}
    \mathbb{D}^L(f)(x)=\sum_{s=1}^{\infty}(-1)^{(s-1)}\frac{f(x+\triangle s)-f(x-\triangle s)}{\triangle s}.\nonumber
\end{align*}
Thus $\mathbb{D}(f)(x)=\lim_{L\rightarrow\infty}\mathbb{D}^L(f)(x)$.

In this section we are restricted to fixed but large values of $L$, and for a given function, we consider the calculation to support the conjecture, $\mathbb{D}(f)(x)= f'(x)$, if for a chosen accuracy goal, $\mathbb{D}^L(f)(x)\approx f'(x)$ to that accuracy.

\paragraph{Shifted Gaussian}
Let us start by considering a shifted Gaussian
\begin{align*}
f(x)=\sqrt{\frac{\pi}{\gamma}}e^{-\gamma(x-x_0)^2},
\end{align*}
 which is covered by Theorem 1.
For the calculations presented in this section we have used the parameter values $\gamma=1$ and $x_0=20$ for the Gaussian function. We evaluate $\mathbb{D}^L(f)$ at the point $\bar{x}=18$, and for the four first figures we use $L=10^5$. We terminate the alternating infinite series defining $\mathbb{D}^L$ for a $s=s_m$ such that the truncation error for the series is less than a prescribed accuracy goal for $\mathbb{D}^L(f)(\bar{x})$. Given this,  $s_m$ is a function of $L$. Since the series is now finite, the calculation of $\mathbb{D}^L(f)(\bar{x})$ only uses points on the x-axis inside the interval $[\bar{x}-s_m\triangle,\bar{x}+s_m\triangle]$.  This is the {\it sample interval} for the calculation. The accuracy goal for the calculations in this section is set to $10^{-8}$.

\begin{figure}[h]
    \centering
    \includegraphics[width=0.8\linewidth]{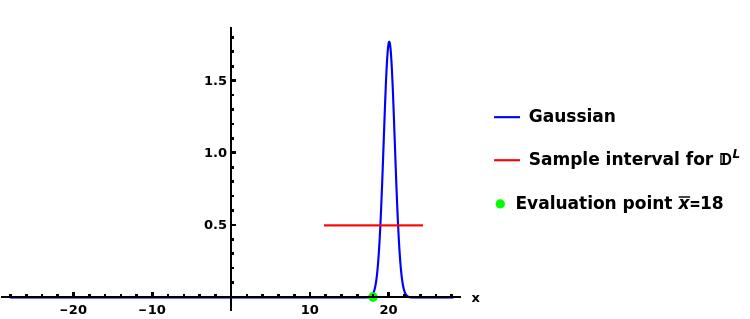}
    \caption{Gaussian function together with  evaluation point $\bar{x}$, and sample interval used for calculating $\mathbb{D}^L(f)(\bar{x})$. $L=10^5$}
    \label{fig5.1.1.1}
\end{figure}

In figure \ref{fig5.1.1.1}, we show the Gaussian function, the point where $\mathbb{D}^L(f)$ is evaluated, and the sample interval for the choice $L=10^5$. The figure clearly illustrates the non-locality of the $\mathbb{D}$ operator.

\begin{figure}[h]
    \centering
    \includegraphics[width=0.8\linewidth]{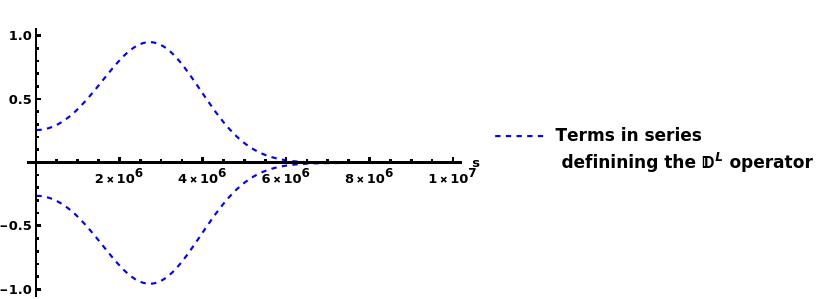}
    \caption{Sequence of terms in the infinite series defining the quantity $\mathbb{D}^L(f)(\bar{x})$ operator.  $L=10^5$, accuracy goal $10^{-8}$.}
    \label{fig5.1.1.2}
\end{figure}

In figure \ref{fig5.1.1.2}, we show the terms in the series defining the quantity $\mathbb{D}^L(f)(\bar{x})$. It is worth noting that for  the given value of $L$, and an accuracy goal of $10^{-8}$ we need to include $10^5$ terms in our series. This is easy to handle numerically  for the Gaussian, but for functions that decay slower than a Gaussian, and for larger values of $L$, and higher accuracy, the calculations quickly become very challenging. We will soon see examples of this.

\begin{figure}[h]
        \centering
        \includegraphics[width=0.6\linewidth]{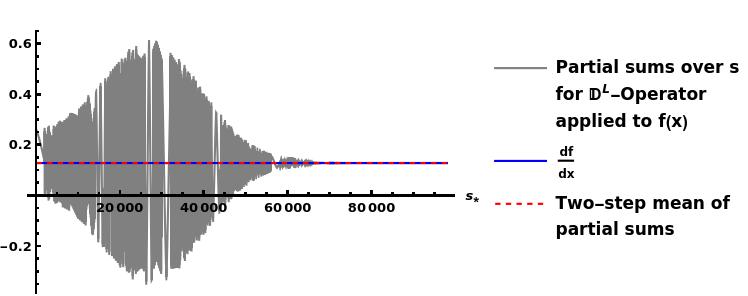}
        \caption{Partial sums of the infinite series defining  the $\mathbb{D}^L$ operator. $L=10^5$}
        \label{fig5.1.1.3}
    \end{figure}

In figure \ref{fig5.1.1.3} we show the partial sums of the series for the quantity $\mathbb{D}^L(f)(\bar{x})$, where we sum the series from $s=1$, to $s=s_*$, for $s^*=1,\dots,s_m$.  If the calculation is to converge, the graph in the figure needs to settle down for increasing values of $s_*$. We see that eventually it does, but that it actually grows for a large stretch of $s_*$ values.  This is typical for what occurs while evaluating the $\mathbb{D}^L$ operator. The growth phase for the partial sums can stretch to very large values of $s_*$, and during this phase, the values of the partial sums can grow extremely large, until they eventually settle down. Handling this growth phase numerically in general requires arbitrary precision arithmetic.
In the figure we also include the exact value for the derivative of $f$ in $x=\bar{x}$. This is the blue horizontal line in the figure. We observe that not only do the partial sums settle down, but they settle down to the exact value for the derivative. Or, to be precise, we find that the difference between $\mathbb{D}^L(f)(\bar{x})$ and $f'(\bar{x})$ when $s_*=s_m$ is less than $10^{-8}$ which was the prescribed precision goal. 

In the figure we also show a curve that represents the sliding mean of each pair of consecutive terms in the partial sums.  This is the red dotted curve in the figure. This curve appears in the figure to coincide with the exact value for the derivative for essentially all the partial sums, starting with $s_*=2$. It seems to tell us that the non-locality of the $\mathbb{D}$ operator is perhaps only apparent;  when evaluating the series determining $\mathbb{D}^L(f)(\bar{x})$, only the first few terms in the sum contribute to the final value, and that this final value comes from the remainder of the  near-cancellation of the first few alternation terms. Consecutive terms, further out in the sum, are the non-local terms that sample values of the Gaussian far from $x=\bar{x}$, cancel up to a remainder that is too small to show up in the plot. This is however not the whole story.

\begin{figure}[ptbh]
        \centering
        \includegraphics[width=0.6\linewidth]{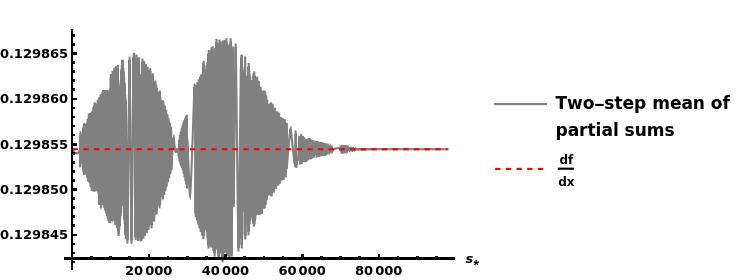}
        \caption{Two-step mean of the partial sums. $L=10^5$}
        \label{fig5.1.1.4}
\end{figure}

In figure \ref{fig5.1.1.4} we leave out the partial sums, retaining only the sliding mean of the partial sums, and the exact value of the derivative, this is now the red dotted curve. Zooming in on these two curves, we see that partial sums which sample values of the Gaussian far from the evaluation point do give a contribution to the value of $\mathbb{D}^L(f)(\bar{x})$. It is small, but these non-local contributions are needed in order to approximate the exact value $f'(\bar{x})$ to the specified accuracy. 
Recall however that according to the conjecture,  $\mathbb{D}^L(f)(\bar{x})$ only needs to coincide with $f'(\bar{x})$ in the limit when $L$ approaches infinity. Thus, the fact that the figures show non-locality for $\mathbb{D}^L$ is not surprising.

\begin{figure}[h!]
    \centering
    \includegraphics[width=0.6\linewidth]{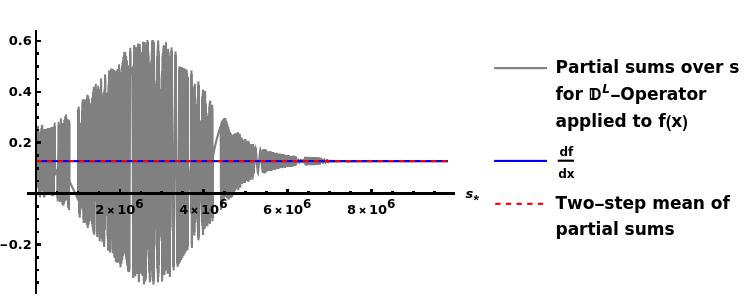}
    \caption{Partial sums of the infinite series defining  the $\mathbb{D}^L$ operator; $L=10^7$}
    \label{fig5.1.1.5}
\end{figure}

\begin{figure}[h!]
    \centering
    \includegraphics[width=0.6\linewidth]{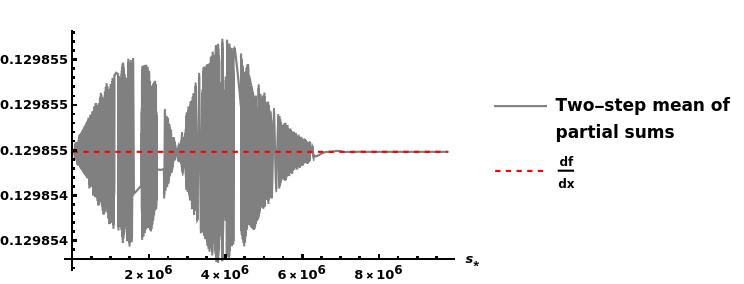}
    \caption{Two-step mean of the partial sums; ; $L=10^7$}
    \label{fig5.1.1.6}
\end{figure}

In figures \ref{fig5.1.1.5} and \ref{fig5.1.1.6} we increase $L$ to the value $L=10^{7}$ and redo the calculations. The curves in the figures look very much the same as before, but in figure \ref{fig5.1.1.6} we see that even if the non-locality appears to be about as extensive as for the smaller value of $L$ in figure \ref{fig5.1.1.4}, the size of the contribution of the non-local terms to the final sum is smaller.

One should of course not expect to be able to calculate $f'(\bar{x})$ to arbitrary precision for fixed $L$ by including enough terms in the series defining $\mathbb{D}^L$. 

Notwithstanding that, for a function as nice as a Gaussian we can for our chosen value of $L=10^5$, squeeze the difference between $\mathbb{D}^L(f)(\bar{x})$ and $f'(\bar{x})$ down to at most $10^{-996}$, and no more. This accuracy is achieved by summing up $10^6$ terms in the sum. In order to reach this accuracy we did the calculations with  a 1000 digit accuracy. This is of course a very high accuracy, and that it is achievable for the Gaussian after only summing up $10^{6}$ terms is most certainly dependent on the fact that the Gaussian decays exponentially away from its center point $x_0$. For functions that decay more slowly, like the Lorentzian function which we will discuss next, to find the achievable accuracy, for a fixed value of $L$, requires  many more terms than for the Gaussian.

\paragraph{Lorentzian}
For the Lorentzian 
\begin{align*}
    f(x)=\frac{1}{1+x^2},
\end{align*}
 which is covered by Theorem 1, we pick the point where we want to evaluate $\mathbb{D}^L(f)$ to be $\bar{x}=2$, and we let, as for the Gaussian, $L=10^5$. Furthermore, we seek the same  accuracy goal as for the Gaussian, $10^{-8}$.

\begin{figure}[h!]
    \centering
    \includegraphics[width=0.6\linewidth]{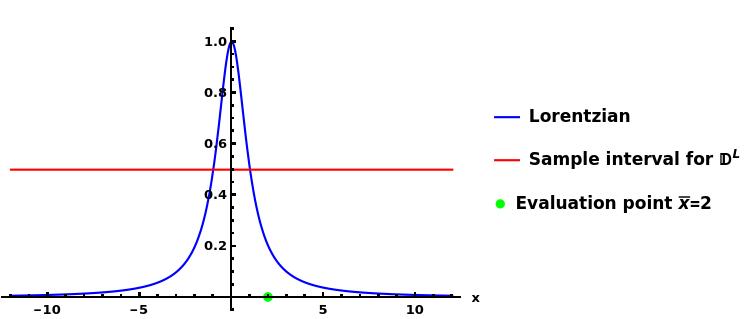}
    \caption{Lorentzian function together with  evaluation point $\bar{x}$, and part of the sample interval used for calculating $\mathbb{D}^L(f)(\bar{x})$. $L=10^5$}
    \label{fig5.1.2.1}
\end{figure}

In figure \ref{fig5.1.2.1} we show the Lorentzian, the evaluation point, and part of the sample interval. The actual sample interval was much larger and approximately given by $[-60,64]$.

\begin{figure}[h!]
    \centering
    \includegraphics[width=0.6\linewidth]{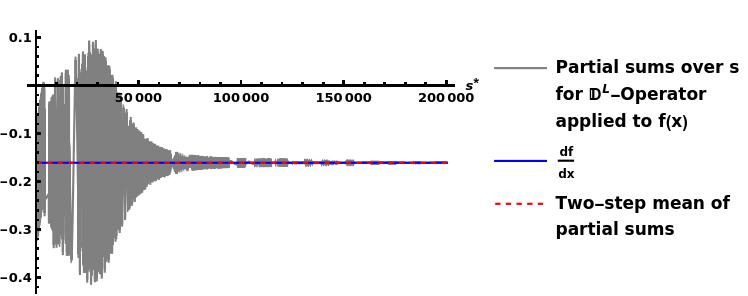}
    \caption{Partial sums of the infinite series defining  the $\mathbb{D}^L$ operator. $L=10^5$}
    \label{fig5.1.2.2}
\end{figure}

\begin{figure}[h!]
    \centering
    \includegraphics[width=0.6\linewidth]{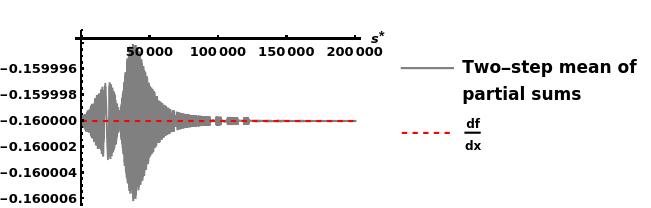}
    \caption{Two-step mean of the partial sums. $L=10^5$}
    \label{fig5.1.2.3}
\end{figure}

The next two figures, \ref{fig5.1.2.2} and \ref{fig5.1.2.3}, correspond to the figures, \ref{fig5.1.1.5} and \ref{fig5.1.1.6} for the Gaussian case. We observe that the quantity $\mathbb{D}^L(f)(\bar{x})$ behaves qualitatively very much like for the Gaussian case.

For our fixed value of $L$, adding more terms in the series keeps increasing the accuracy, as in the Gaussian case,  but for the Lorentzian the accuracy levels off at  $10^{-1921}$ after summing $10^{550}$ terms. 

Note that to achieve this kind of accuracy, we are not actually summing up the required $10^{550}$ terms, but rather are relying on well known methods that estimate the sum of alternating sums, without actually computing all the terms\cite{Wolfram}.

\paragraph{Tent function}\label{TheTentFunction}
For the tent function 
\begin{align*}
    f(x)=\left\{ \begin{aligned}
x+1,&\;\;\; -1\leq x\leq 0&\\
-x+1,&\;\;\;\;\;\;\;\; 0\leq x\leq 1&\\
0,&\;\;\;\;\;\;\;\;\; \abs{x}> 1&\\
\end{aligned}  \right.
    ,
\end{align*}
 which is {\it not} covered by Theorem 1, 
we pick the point where we want to evaluate $\mathbb{D}^L(f)$ to be $\bar{x}=-1/2$, and the same accuracy, $10^{-8}$ as the other examples above. In order to achieve this accuracy for the tent functions we need to increase the value of $L$ to $10^{7}$ and sum up ca. $10^{7}$ terms. Note that this $L$ is two orders of magnitude larger than the values of $L$ we needed for the Gaussian and Lorentzian cases.
\begin{figure}[h!]
    \centering
    \includegraphics[width=0.6\linewidth]{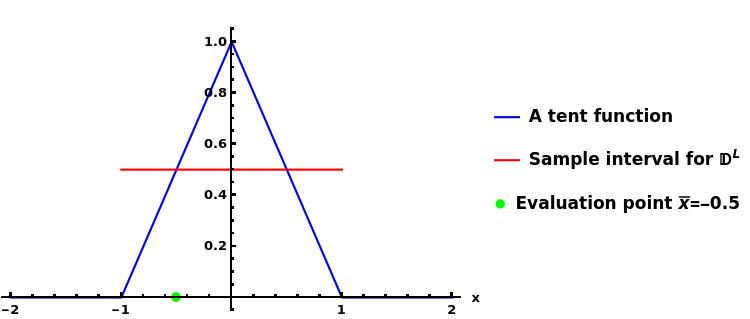}
    \caption{The tent function together with  evaluation point $\bar{x}$, and the sample interval used for calculating $\mathbb{D}^L(f)(\bar{x})$. $L=10^7$}
    \label{fig5.1.4.10}
\end{figure}

Figure \ref{fig5.1.4.10},we shows the tent function, the chosen evaluation point, and the sample interval. Clearly, since the tent function has compact support, the sample interval can never be larger than the support. For the parameters we are using, the sample interval is as large as it can be, $[-1,1]$.

\begin{figure}[h!]
    \centering
    \includegraphics[width=0.6\linewidth]{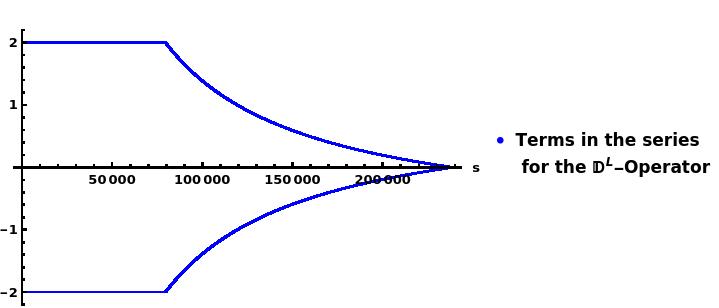}
    \caption{Sequence of terms in the infinite series defining the quantity $\mathbb{D}^L(f)(\bar{x})$, $L=10^7$, accuracy goal $10^{-8}$.}
    \label{fig5.1.4.2}
\end{figure}
\begin{figure}[h]
    \centering
    \includegraphics[width=0.7\linewidth]{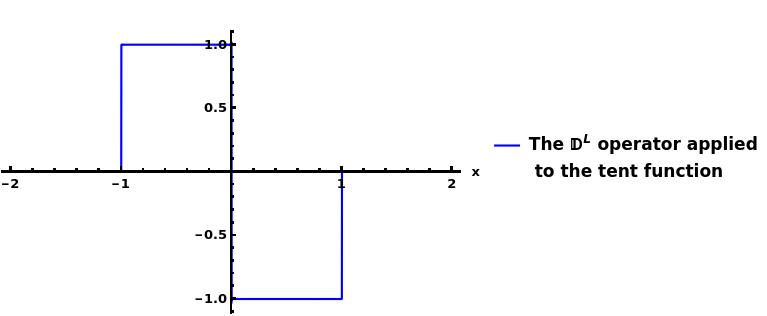}
    \caption{The function $\mathbb{D}^L(f)(x)$}
    \label{fig5.1.4.3}
\end{figure}

In figure \ref{fig5.1.4.2}  show the terms in the sequence defining the quantity $\mathbb{D}^L(f)(\bar{x})$, and in figure \ref{fig5.1.4.3} we show the function $\mathbb{D}^L(f)(x)$. It is evident that $\mathbb{D}^L(f)(x)\approx f'(x)$ for all $x$ where $f(x)$ is differentiable.

Note that even if the tent function is not differentiable at the points ${-1,0,1}$, we can still try to apply the $\mathbb{D}^L$-operator to the tent function at these points. The numerical calculations we have done strongly indicate that 
\begin{align*}
    \mathbb{D}(f)(x)=\frac{1}{2}\left(\lim_{x->x^+}f'(x)+\lim_{x->x^-}f'(x)\right)=\{\frac{1}{2},0,-\frac{1}{2}\},\;\;\;\;\text{for}\;\;\;\;\;\;x\in\{-1,0,1\}.
\end{align*}
Note that for the tent function, like for a few of the following functions in this section, that have compact support and a simple form, we can find an  explicit formula for the key expression $\mathbb{D}^L(f)(x)$.

\paragraph{Hat function}\label{TheHatFunction}
In this subsection we introduce the following {\it hat} function

\begin{align*}
    f(x)=\left\{ \begin{aligned}
0,&\;\;\;\;\;\;\;\; x\leq 0&\\
1,&\;\;\;\;\;\;\;\; 0< x<l&\\
0,&\;\;\;\;\;\;\;\;\; x\geq 0&\\
\end{aligned}  \right.
    ,
\end{align*}
where $l>0$. This function is clearly not covered by Theorem 1.

 For this simple function we can also find an exact formula for the quantity $\mathbb{D}^L(f)(\bar{x})$.
 
If we  let the evaluation point be $\bar{x}=\frac{l}{2}-1$, the distance between the evaluation point and both points of discontinuity increases when $l$ increases.  At this point we have $f'(\bar{x})=0$.

\begin{figure}[h!]
    \centering
    \includegraphics[width=0.7\linewidth]{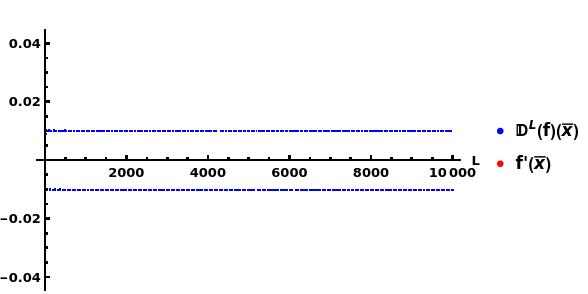}
    \caption{The sequence $\mathbb{D}^L(f)(\frac{l}{2}-1)$ for $l=20$, and increasing $L$ in the range $1\le L\le 10^{4}$.}
    \label{fig5.1.6.1}
\end{figure}

In figure \ref{fig5.1.6.1} we plot the sequence $\mathbb{D}^L(f)(\bar{x})$ for increasing values of $L$ in the range $1\le L\le 10^{3}$. The value of the sequence fluctuates around the correct value of the derivative. The amplitude of the fluctuations are on the order of $10^{-2}$. 

\begin{figure}[h!]
    \centering
    \includegraphics[width=0.7\linewidth]{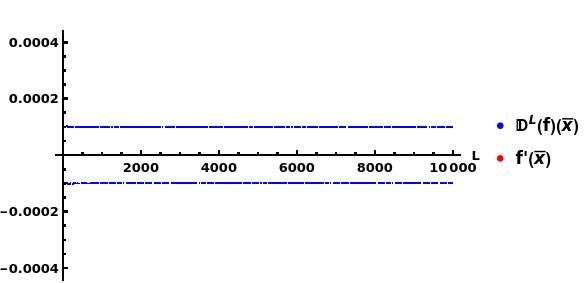}
    \caption{The same as in figure \ref{fig5.1.6.1}, but now $l=200$.}
    \label{fig5.1.6.2}
\end{figure}

In figure \ref{fig5.1.6.2} we increase $l$ to the value $l=200$ while keeping the evaluation point fixed. The evaluation point is now further from the two points of discontinuity than in figure \ref{fig5.1.6.1}. We see that the sequence $\mathbb{D}^L(f)(\bar{x})$ for increasing $L$ fluctuate around the correct value of the derivative, but the amplitude of the fluctuations are now of order $10^{-4}$.

These two figures tell us that as the distance between where we want to compute $\mathbb{D}(f)(\bar{x})$, and the point of discontinuity, increases, the fluctuating value of $\mathbb{D}^L(f)(\bar{x})$ gets ever closer to the exact value of the derivative $f'(\bar{x})$. However, as far as we can tell from our numerics, no matter how far away from $x=\bar{x}$ the point of discontinuity is, the fluctuating values of $\mathbb{D}^L(f)(\bar{x})$ {\it never} settle down, and thus $\mathbb{D}(f)(\bar{x})$ does not exist. Thus, this numerical example indicate that for the hat function, our conjecture does not hold for all $\bar{x}$.

From the formula it is easy to verify that for the symmetrically placed point $\bar{x}=\frac{l}{2}$, we find
$\mathbb{D}(f)(\bar{x})=f'(\bar{x})=0$. Thus, for this symmetrically placed point the conjecture does holds.

\paragraph{Truncated ramp function}
Here we extend our exploration of the $\mathbb{D}$-operator  for the case of discontinuous functions by introducing the following {\it truncated ramp} function, which evidently is not covered by Theorem 1,
\begin{align*}
    f(x)=\left\{ \begin{aligned}
0,&\;\;\;\;\;\;\;\; x\leq -1&\\
x+1,&\;\;\; -1\leq x\leq 0&\\
1-\frac{1-h}{l}x,&\;\;\;\;\;\;\;\; 0\leq x < l&\\
0,&\;\;\;\;\;\;\;\;\; x\ge l&\\
\end{aligned}  \right.
    ,
\end{align*}
The function is continuous everywhere, except at the point $x=l$, and is continuously differentiable everywhere else, except for $x=-1,0,l$. The size of the drop at the point of discontinuity is given by $h$.

\begin{figure}[h!]
    \centering
    \includegraphics[width=0.7\linewidth]{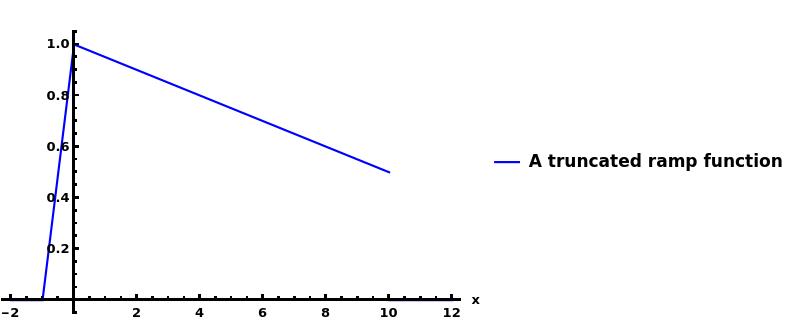}
    \caption{The truncated ramp function for $h=0.5$, and $l=10$.}
    \label{fig5.1.5.1}
\end{figure}
\begin{figure}[h!]
    \centering
    \includegraphics[width=0.7\linewidth]{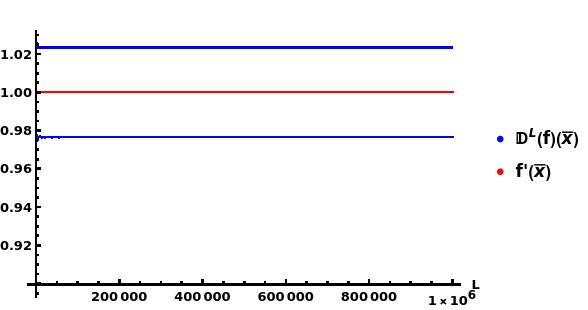}
    \caption{$\mathbb{D}^L(f)(\bar{x})$ for increasing value of $L$,together with the derivative of the truncated ramp function for the case $\bar{x}=-0.67$, $h=0.5$, and $l=10$}
    \label{fig5.1.5.2}
\end{figure}
\begin{figure}[h!]
    \centering
    \includegraphics[width=0.7\linewidth]{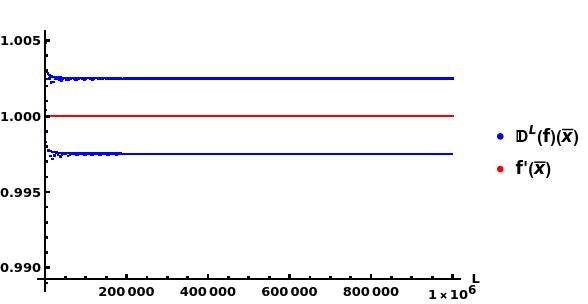}
    \caption{Same as figure \ref{fig5.1.6.2}, but now for $l=100$}
    \label{fig5.1.5.3}
\end{figure}

Figure \ref{fig5.1.5.1} is a graph of the truncated ramp function for the case $l=10$, and $h=0.5$.
The function has compact support, and a formula for sum of the series determining $\mathbb{D}^L(f)(\bar{x})$ can be found. In figures \ref{fig5.1.5.2} and \ref{fig5.1.5.3} we use this formula to plot, for increasing value of $L$,  $\mathbb{D}^L(f)(\bar{x})$, with $\bar{x}=-0.67$. At this point we have  $f'(\bar{x})=1$.

These two figures tell us that as the distance between where we want to compute $\mathbb{D}(f)(\bar{x})$, and the point of discontinuity, increases, the fluctuating value of $\mathbb{D}^L(f)(\bar{x})$ gets ever closer to the exact value of the derivative $f'(\bar{x})$. However, as far as we can tell from our numerics, no matter how far away from $x=\bar{x}$ the point of discontinuity is, the fluctuating values of $\mathbb{D}^L(f)(\bar{x})$ {\it never} settle down, and thus $\mathbb{D}(f)(\bar{x})$ does not exist. 

Keeping $l$ fixed, but reducing the size of the drop at the point of discontinuity leads to the same behavior for the fluctuating values of
$\mathbb{D}^L(f)(\bar{x})$. Thus no matter how small the drop at the discontinuity is, $\mathbb{D}(f)(\bar{x})$ does not exist.

Based on this and more numerical calculations, for other discontinuous functions, that we don't show here, we conjecture that we will get similar results for any function that is discontinuous at one or more points. Thus, it does not matter how smooth it is away from the points of discontinuity, the mere presence of a discontinuity, anywhere, will in general invalidate our original conjecture (\ref{conjecture}), except perhaps for some special symmetrically placed points where the influence from the points of discontinuity cancel. 

One could say that the existence of discontinuities {\it unmask} the apparent locality of the $\mathbb{D}$-operator we have seen while applying it to smooth functions.

\paragraph{Square root}
For the square root  
\begin{align*}
    f(x)=\sqrt{x},\;\;x>0,
\end{align*}
which is not covered by Theorem 1,
we pick the point where we want to evaluate $\mathbb{D}^L(f)$ to be $\bar{x}=1$, and we let the value of $L$, and the accuracy be the same as for the Gaussian and the Lorentzian.
\begin{figure}[h!]
    \centering
    \includegraphics[width=0.7\linewidth]{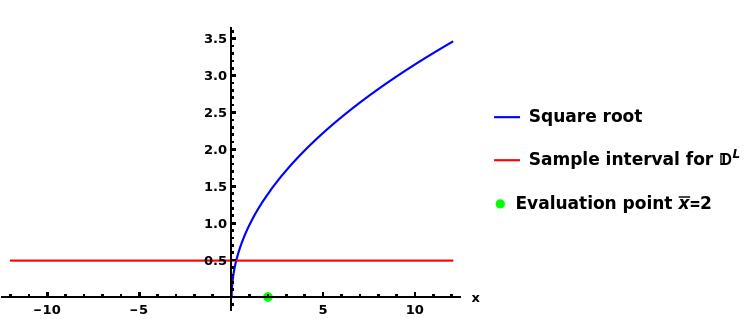}
    \caption{Square root function together with  evaluation point $\bar{x}$, and part of the sample interval used for calculating $\mathbb{D}^L(f)(\bar{x})$. $L=10^5$}
    \label{fig5.1.3.1}
\end{figure}
In figure \ref{fig5.1.3.1} we show the square root, the evaluation point, and part of the sample interval. Like for the Lorentzian, the actual sample interval for the square root is much larger than shown. We find that for the accuracy we ask for, $10^{-8}$, the number of terms  the series defining the quantity $\mathbb{D}^L(f)(\bar{x})$ is much too large to sum up directly. For example, summing up $10^{10}$ terms only give us the accuracy $10^{-2}$. In order to achieve the desired accuracy of $10^{-8}$, we must use the estimation methods like we did for the Lorentzian. If we do this we can achieve our desired accuracy of $10^{-8}$. The sampling interval we need for the square root in order to achieve this accuracy is enormous, and very much larger than for the Gaussian and the Lorentzian. This must surely be connected to the fact that the square root doesn't decrease at all when $x$ grows, it actually increases.

Summing the series for the quantity $\mathbb{D}^L(f)(\bar{x})$, using the estimation method, we find that we need ca. $10^{20}$ terms in order to achieve an accuracy of $10^{-8}$.
With respect to the achievable accuracy for the assumed value of $L$, we found that it is at least $10^{-55}$.

There is however an issue that arises while calculating  $\mathbb{D}^L(f)(\bar{x})$ for the square root; the sample interval include negative $x$-values already for an accuracy of $10^{-8}$. For the computations we are doing in this paper, the square root function is defined to have a branch cut on the negative real axis. The values of the square root along the negative axis takes on purely imaginary values. So how come the quantity $\mathbb{D}^L(f)(\bar{x})$ is real? The only way this can occur is if all the imaginary parts in the series for the $\mathbb{D}^L$-operator cancel, up to the accuracy of the calculation. Computational investigations verify that this is indeed so. However, we don't have a transparent mathematical explanation for why cancellation of all imaginary quantities takes place.

\paragraph{Complex functions}
In this section we generalize the $ \mathbb{D}$ operator to an operator on functions, $f(z)$,  of a complex variable. The formula for the operator is
\begin{align*}
 \mathbb{D}_{\xi}(f)(z)=\lim_{L\rightarrow\infty}\sum_{s=1}^{\infty}(-1)^{(s-1)}\frac{f(z+\xi\triangle s)-f(z-\xi\triangle s)}{\xi\triangle s},
\end{align*}
where $\xi$ is a complex number of unit length, $\abs{\xi}=1$.  Thus for the operator $ \mathbb{D}_{\xi}$ our sampling occur along a straight line in the complex plane. The direction of the line is determined by the complex number $\xi$.

\begin{figure}[h!]
    \centering
    \includegraphics[width=0.7\linewidth]{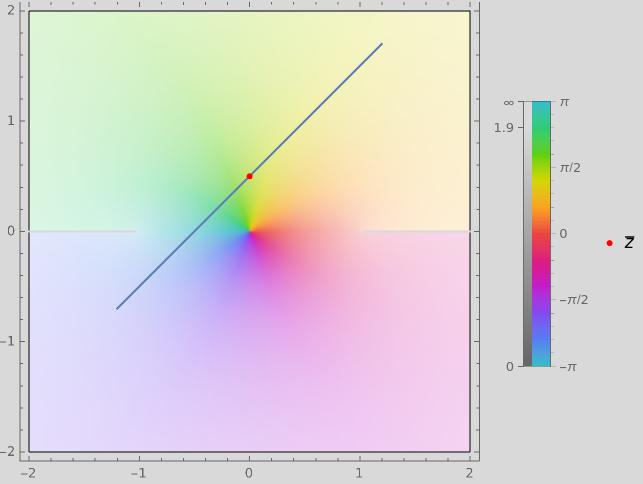}
    \caption{Plot of  argument of the $\arcsin$, together with the evaluation point $\bar{z}$ and part of the sample line. The branch cuts are displayed in gray.}
    \label{fig5.1.4.1}
\end{figure}

Let us consider the complex function
\begin{align*}
    f(z)=\arcsin(z).
\end{align*}
Let us pick the point where we want to evaluate $ \mathbb{D}_{\xi}(f)$ to be $\bar{z}=0.5i$,  let $L=10^{4}$, the accuracy be $10^{-6}$, and let $\xi=\frac{1}{\sqrt{2}}(1+i)$.
In figure \ref{fig5.1.4.1} we show a plot of the complex argument of $f(z)$, together with the evaluation point and a small section of the sample line. The branch cuts for the function are marked with gray.

We find that in order to achieve the desired accuracy we need to include $10^{10}$ terms in the series defining the quantity $ \mathbb{D}_{\xi}(f)(\bar{z})$. This is the highest accuracy we can reach in a reasonable time while directly summing up the terms in the series. Summing up the series using estimation methods, we find that using $10^{20}$ terms in the series, and calculating with $25$ digits accuracy, $ \mathbb{D}_{\xi}(f)(\bar{z})=f'(z)$ within machine precision. 

We have also observed using numerical computations that the quantity $ \mathbb{D}_{\xi}(f)(\bar{z})$ does not appear to depend on $\xi$, meaning that the quantity is independent on which sample line, through the point $\bar{z}$, we use. Since $f(z)=\arcsin(z)$ is analytic at the point $\bar{z}=0.5i$, this is perhaps not so unexpected. If our $ \mathbb{D}_{\xi}$ operator is equal to the complex derivative, this is of course exactly what would have to happen.  On the same theme, we find that if we apply our operator to a complex function that is {\it not} analytic, for example $f(z)=\abs{z}$, then the quantity $ \mathbb{D}_{\xi}(f)(z)$ {\it does} depend on the direction of the sample line.

We have done investigations like the above for several different functions of a complex variable, and all these cases support the complex version of the original conjecture (\ref{conjecture}), 
\begin{align*}
     \mathbb{D}_{\xi}(f)(\bar{z})=\frac{df}{dz}(\bar{z}),
\end{align*}
independently of $\xi$, if $f$ is analytic in $z=\bar{z}$.

Our original conjecture, which was inspired by deriving the continuum limit of the total angular momentum of a scalar field in a square cavity, was restricted to functions defined on the real axis. The sample line was also restricted to being located on the real axis. However, if a function on the real axis can be extended to complex variables, one can apply the extended $ \mathbb{D}_{\xi}$ operator discussed in this section to compute $ \mathbb{D}_{\xi}(f)(x)$ for x on the real line, where the sample line now is in the complex plane. We have investigated several such situations where the extended function is analytic in a domain containing the real axis. In all such cases we find that 
\begin{align*}
     \mathbb{D}_{\xi}(f)(x)= \mathbb{D}(f)(x),
\end{align*}
independently of $\xi$. Thus, we conjecture that $ \mathbb{D}_{\xi}$ is truly an extension of $\mathbb{D}$.


\bigskip

\begin{thebibliography}{10}
\bibitem{Jones}
D. S. Jones, "The theory of electromagnetism", page 14,  Pergamon Press, (1964).
\bibitem{Cohen}
C. Cohen-Tannoudji, J. Dupont-Roc, and G. Grynberg, {\it Photons and Atoms: Introduction to Quantum Electrodynamics}, Wiley, New York (1989). See complement BI, Eq.(11).
\bibitem{Haus}
H. A. Haus and J. L. Pan, “Photon spin and the paraxial wave equation,” Am. J. Phys. 61, 818-821 (1993).
\bibitem{Wolfram}
Wolfram Research (1988), NSum, Wolfram Language function, https://reference.wolfram.com/language/ref/NSum.html (updated 2007).
\bibitem{Weinberg}
S. Weinberg, "The quantum theory of fields", Volume 1, page 306, Cambridge, (1995).
\bibitem{Jackson}
J.D. Jackson,{\it  Classical Electrodynamics} (3rd edition), Wiley, New York (1999).
\bibitem{Barnett1}
S. M. Barnett, “Optical angular-momentum flux,” J. Opt. B: Quantum Semiclassical Optics 4, S7-16 (2002).
\bibitem{Allen}
L. Allen, S. M. Barnett, and M. J. Padgett," Optical Angular Momentum", Institute of Physics Publishing, Bristol, United Kingdom (2003).
\bibitem{Grynberg}
G. Grynberg, A. Aspect, and C. Fabre,{\it Introduction to Quantum Optics}, Cambridge University Press, Cambridge, United Kingdom (2010).
\bibitem{Barnett2}
S. M. Barnett, “Rotation of electromagnetic fields and the nature of optical angular momentum,” J. Mod. Opt. 57, 1339-1343 (2010).
\bibitem{Masud1}
M. Mansuripur,“Spin and orbital angular momenta of electromagnetic waves in free space,”{\it Physical Review A} 84, 033838 (2011)
\bibitem{Marrucci}
L. Marrucci, E. Karimi, S. Slussarenko, B. Piccirillo, E. Santamato, E. Nagali, and F. Sciarrino, “Spin-to-orbital conversion of the angular momentum of light and its classical and quantum applications,” J. Opt. 13, 064001 (2011).
\bibitem{Per}
P. K.  Jakobsen, "Topics in applied Mathematics and nonlinear waves", (2019), doi.org/10.48550/arXiv.1904.07702.
\bibitem{Masud2}
M. Mansuripur, “Spin and orbital angular momenta of electromagnetic waves in classical and quantum electro-dynamics,” {\it Proceedings of SPIE} 12656, Spintronics XVI 1265604 (2023); doi: 10.1117/12.2675779.
\bibitem{Masud3}
M. Mansuripur, “Electromagnetic angular momentum of quantized wavepackets in free space,” {\it Proceedings of SPIE }12649, Optical Trapping and Optical Micromanipulation XX, 126490A (2023); doi: 10.1117/12.2675775
\end{thebibliography}
\end{document}